\documentstyle[aps,floats,epsfig]{revtex}
\begin{document}
\draft

\title{Evolution of Parton Fragmentation Functions at Finite Temperature}

\author{Jonathan A. Osborne$^1$, Enke Wang$^{2,1}$ and Xin-Nian Wang$^{1,3}$}

\address{$^1$Nuclear Science Division, Lawrence Berkeley Laboratory,
         Berkeley, California 94720, USA}
\address{$^2$Institute of Particle Physics, Huazhong Normal University,
         Wuhan 430079, China}
\address{$^3$Department of Physics, Shandong University,
         Jinan 250100, China}

\date{October 25, 2002}

\maketitle

\begin{abstract}
The first order correction to the parton fragmentation functions 
in a thermal medium is derived in the leading logarithmic approximation
in the framework of thermal field theory. The medium-modified evolution 
equations of the parton fragmentation functions are also derived.
It is shown that all infrared divergences, both linear and logarithmic,
in the real processes are canceled among themselves and by corresponding 
virtual corrections. The evolution of the quark number and the energy 
loss (or gain) induced by the thermal medium are investigated.
\end{abstract}



\section{Introduction}
\label{sec1}

Parton production in hard processes is normally followed by final-state
radiation and subsequent hadronization, giving rise to collimated jets
of hadrons as observed in experiments. The hadron distributions inside a
jet known as jet fragmentation functions can be defined as the 
vacuum expectation values of the parton
fields and the hadronic interpolating fields \cite{CSS89}. 
Though these fragmentation functions are non-perturbative and currently 
can only be measured experimentally, their evolution with the probing 
scale can be calculated within perturbative QCD. The resultant
Dokshitzer-Gribov-Lipatov-Altarelli-Parisi (DGLAP) \cite{dglap} evolution 
equations have been stringently tested against experiments and now can 
even be used to measure the scale-dependence of the running strong 
coupling constant \cite{bethke}.
If the hard parton is produced amid a thermal QCD medium, as
most likely occurs in high-energy heavy-ion collisions, the
subsequent final-state radiation and parton cascade must be modified by
the presence of the medium. One not only has to consider parton emission
but also parton absorption \cite{ww01} in describing the evolution of the
fragmentation functions. Such a medium effect is closely related to
the radiative parton energy loss, which also leads to modified fragmentation
functions \cite{whs}. In most of the studies of parton energy loss, one
relies on a Debye-screened potential model \cite{gw94,dmps,glv,wied} for
parton-medium interaction. One then has to introduce additional parameters
such as the mean free path, which is ill-defined when the magnetic
part of the one-gluon exchange interaction is included. One can replace
the mean free path by a transport parameter \cite{dmps} and eliminate
the infrared problem by using a hard-thermal-loop (HTL) resummed gluon 
propagator \cite{wang00}. However, it is still theoretically interesting
to study the problem within the framework of field theory at finite
temperature.

This paper is our first attempt to study medium-modified fragmentation
functions within QCD field theory at finite temperature. We will first
extend the definition of the fragmentation functions to include the scenario
of parton propagation inside a thermal bath. We will calculate
the first order corrections to the parton fragmentation functions in a 
thermal medium in the leading logarithmic approximation and derive
the corresponding modified evolution equations. We will find that the
modified evolution equation can be cast into a similar form as
the DGLAP equations in the vacuum. However, the modified splitting
functions depend explicitly on the temperature and the partons' 
initial energy. We will study the structure of different contributions
and their physical interpretations. We will demonstrate that all infrared
(both linear and logarithmic) divergences in radiative corrections 
cancel either among themselves or
with the virtual corrections. We will also study the evolution of the net
quark number and energy loss (or gain).

We should emphasize that our calculation in this paper includes only
the first order corrections in the leading logarithmic approximation. At
this order, we can only consider parton emission and absorption. By
solving the evolution equations, one can effectively resum radiative
corrections associated with the leading parton. However, collision-induced 
radiation is not considered at the first order of perturbation. 
Furthermore, we cannot include any interference effect 
like the Landau-Pomeranchuck-Migdal (LPM) 
interference \cite{lpm} in medium-induced bremsstrahlung 
in the leading log approximation.
One has to include the spectral function of 
the HTL resummed gluon propagator and go beyond the 
leading log approximationin order to consider 
radiation induced by multiple scattering and the LPM
interference effect. We leave this to future investigation.

The remainder of the paper is organized as follows. In Sec. II, we will
first review the definitions of the fragmentation functions in the vacuum
and the basic physical processes that lead to the DGLAP evolution equations
of the fragmentation functions. We then extend the definition to include 
the case in which a parton propagates 
through a thermal QCD medium. In Sec. III, we derive the 
radiative corrections to the fragmentation functions in a thermal medium
based on finite temperature field theory. We will discuss various
physical processes that contribute to the radiative corrections and
the cancellation of all infrared divergences.
In Sec. IV, the QCD evolution equations are derived from the
calculated radiative corrections. As examples of the application of
the evolution equations, we derive the evolution equations for the
net quark number and calculate the energy loss (gain) induced by 
the medium. Finally, we give a summary and conclusion in Sec. V.

\section{Fragmentation Functions at Zero and Finite Temperature}
\label{sec2}

\subsection{Fragmentation Function at Zero Temperature}
\label{sec2a}

Semi-inclusive cross sections can generally be 
factorized into the convolution of 
a parton fragmentation function and a hard partonic cross
section in the collinear (leading twist) approximation. 
Fragmentation functions are
defined as the vacuum expectation values of parton fields and
hadronic interpolating fields. 
The $e^+e^-$ annihilation process is an ideal 
framework in which to define the quark and anti-quark fragmentation
functions. The total hadronic cross section for this process can be
expressed as
 \begin{equation}
 \label{crosssec}
  \sigma_{e^+e^-\rightarrow X}=\frac{2\pi}{s}\frac{e^4}{q^4}
  L_{\mu\nu}(p_a,p_b)W^{\mu\nu}(q)\, ,
 \end{equation}
where $s=(p_a+p_b)^2$ is the invariant mass of electron-positron
system and $q=p_a+p_b$ the momentum of virtual photon $(\gamma^*)$.
The leptonic tensor $L_{\mu\nu}$ is given by
 \begin{equation}
 \label{lepton}
  L_{\mu\nu}(p_a,p_b)=\frac{1}{4}{\rm Tr}\left(\gamma_{\mu}{\not p}_a
    \gamma_{\nu}{\not p}_b\right)\, ,
 \end{equation}
where $1/4$ comes from the average over spin polarization of the initial
$e^+e^-$ state. The hadronic tensor is defined as
 \begin{eqnarray}
 \label{hadron1}
  W^{\mu\nu}(q)&=&\frac{1}{4\pi}\sum_{X}
  \langle 0|J^{\mu}(0)|X\rangle\langle X|J^{\nu}(0)|0\rangle
(2\pi)^4\delta^4(p_{_X}-q)
 \nonumber\\
   &=&\frac{1}{4\pi}\int d^4x e^{-i q\cdot x}
   \langle 0|J^{\mu}(0)J^{\nu}(x)|0\rangle\, , \label{eq:jj}
 \end{eqnarray}
where $\sum_X$ runs over all possible intermediate states and the
quark electromagnetic current is $J_{\mu}=\sum_q Q_q{\bar
\psi}_q\gamma_{\mu}\psi_q$.  Here, $Q_q$ is the electric charge of
the quark in units of the proton charge.

At the lowest order in pQCD, the electron-positron pair annihilates, 
forming a virtual photon which then decays into a
quark anti-quark pair.  After a sufficiently long time, 
the partons undergo a nonperturbative fragmentation 
process and emerge from the scattering center 
in the form of hadronic jets.  
The two partons fragment independently of one another
in the leading twist approximation, allowing us to define
process-independent fragmentation functions.
In terms of these fragmentation functions, 
the semi-inclusive cross section $\sigma_{e^+e^-\rightarrow h}$
can be shown to have the form
 \begin{equation}
 \label{inclusec}
  \frac{d\sigma_{e^+e^-\rightarrow h}}{dz_h}
    =\sum_q\sigma_{0}^{q\bar q}
    \left[D_{\bar q \rightarrow  h}(z_h)
    +D_{q\rightarrow h}(z_h)\right]
 \end{equation}
in the center-of-mass frame.
Here, 
\begin{equation}
\sigma_{0}^{q\bar q}\equiv N_c\frac{4\pi\alpha^2 Q^2_q}{3s}
\end{equation}
is the perturbative cross section in the lowest order
for $e^+e^-\rightarrow q\bar q$
and $D_{q(\bar q)\rightarrow h}(z_h)$ is the probability that
a quark (anti-quark) will decay into a hadron $h$ with a fraction
$z_h$ of its energy.  $N_c$ is the number of quark colors.
The energy of the quark (anti-quark) 
is just $E_{q(\bar q)}=\sqrt{s}/2$, so the observed hadron 
has energy $p_h^0=z_h\sqrt{s}/2$.  This relationship 
can also be written in the invariant form 
\begin{equation}
z_h={2p_h\cdot q\over q^2}\;\; .
\end{equation}
Applying the collinear approximation to Eq.~(\ref{eq:jj}),
one can obtain a formal definition
of the fragmentation functions \cite{CSS89}:
 \begin{eqnarray}
 \label{qfrag}
  D_{q({\bar q})\rightarrow h}(z_h)
  &=&\frac{z_h^3}{2}T_{q({\bar q})}(z_h)
 \nonumber\\
  &=&\frac{z_h^3}{4}\int\frac{d^4 k}{(2\pi)^4}
   \delta\left(z_h {-} \frac{p_h\cdot n}{k\cdot n}\right)
   {\rm Tr}\left\lbrack\frac{\gamma\cdot n}{p_h\cdot n}
   {\hat T}_{q({\bar q})}(k, p_h)\right\rbrack\, ,
 \end{eqnarray}
where the Dirac operators ${\hat T}_{q(\bar q)}(k,p_h)$ are given by
 \begin{mathletters}
 \label{Tqmatrix}
 \begin{eqnarray}
   ({\hat T}_{q})_{\alpha\beta}(k, p_h)
 &=&\int d^4 xe^{-ik\cdot x}
    \sum_{S}\langle 0|{\psi}_\alpha(0)|p_h,S\rangle
    \langle p_h,S|\bar\psi_\beta(x)|0\rangle\, ,
 \label{Tqmatrix1}\\
 ({\hat T}_{\bar q})_{\alpha\beta}(k, p_h)
&=&\int d^4 xe^{-ik\cdot x}
    \sum_{S}\langle 0|\bar\psi_\beta(0)|p_h,S\rangle
    \langle p_h,S|{\psi}_\alpha(x)|0\rangle\, .
 \label{Tqmatrix2}
 \end{eqnarray}
 \end{mathletters}
\noindent Here, the sums go over all physical states $S$
and sums over quark colors are implicitly averaged.
The light-like vector $n^\mu\equiv [n^+, n^-, n_{\perp}]
=[0, 1,0_\perp]$ is taken conjugate to the 
momentum of the observed hadron in the sense that
its spatial components are antiparallel to the 
spatial momentum of the hadron.  This implies
$n\cdot p_h=p_h^+=(p_h^0+|\vec p_h|)/2$.
The gauge links required to make this expression
gauge-invariant have been suppressed since 
they do not contribute to the leading-twist fragmentation
functions in the light-cone gauge, $n\cdot A=0$.

The gluon fragmentation function $D_{g\rightarrow
h}(z_h)$ are defined in a similar way:
 \begin{eqnarray}
 \label{gfrag1}
 D_{g\rightarrow h}(z_h)
   &=&\frac{z_h}{2}T_g(z_h)
 \nonumber\\
&=&\frac{z_h^2}{2}\int\frac{d^4 k}{(2\pi)^4}
   \delta\left(z_h-\frac{p_h\cdot n}{k\cdot n}\right)
   d_{\mu\nu}(k)
   {\hat T}_g^{\mu\nu}(k, p_h)\, ,
 \end{eqnarray}
with
 \begin{eqnarray}
 \label{Tgmatrix}
{\hat T}_g^{\mu\nu}(k, p_h)
   &=&\int d^4 xe^{-ik\cdot x}\sum_{S}
    \langle 0|A^{\mu}(0)|p_h,S\rangle
    \langle p_h,S|A^{\nu}(x)|0\rangle\, .
 \end{eqnarray}
\noindent Similarly, gluonic color indices are 
implicitly averaged over.
The gluon polarization vectors $\varepsilon^\mu(k)$ satisfy
 \begin{eqnarray}
 \label{polar}
   d_{\mu\nu}(k)&\equiv&\sum_{\lambda=1,2}
   \varepsilon_{\mu}(k,\lambda)\varepsilon_{\nu}(k,\lambda)
 \nonumber\\
   &=&-g_{\mu\nu}+\frac{k_{\mu}n_{\nu}+k_{\nu}n_{\mu}}{n\cdot k}
 \end{eqnarray}
in the light-cone gauge.
The field-strength tensor, 
$G_{\mu\nu}=
\partial_{\mu}A_{\nu}-\partial_{\nu}A_{\mu}+ig[A_\mu,A_\nu]$,
allows us to write the gluon fragmentation function
(\ref{gfrag1}) in the manifestly gauge-invariant form
 \begin{equation}
 \label{gfrag2}
  D_{g\rightarrow h}(z_h)=-\frac{z_h^2}{2p_h^+}\sum_{S}
   \int\frac{dx^-}{2\pi}e^{-ip_h^+ x^-/z_h}
\langle 0|G^{+\mu}(0)|p_h,S\rangle
    \langle p_h,S|{G^+}_{\mu}(x^-)|0\rangle \, .
\end{equation}
Once again, the gauge links necessary to make this expression
gauge-invariant do not contribute to our functions
in the light-cone gauge.

\begin{figure}
\centerline{\psfig{figure=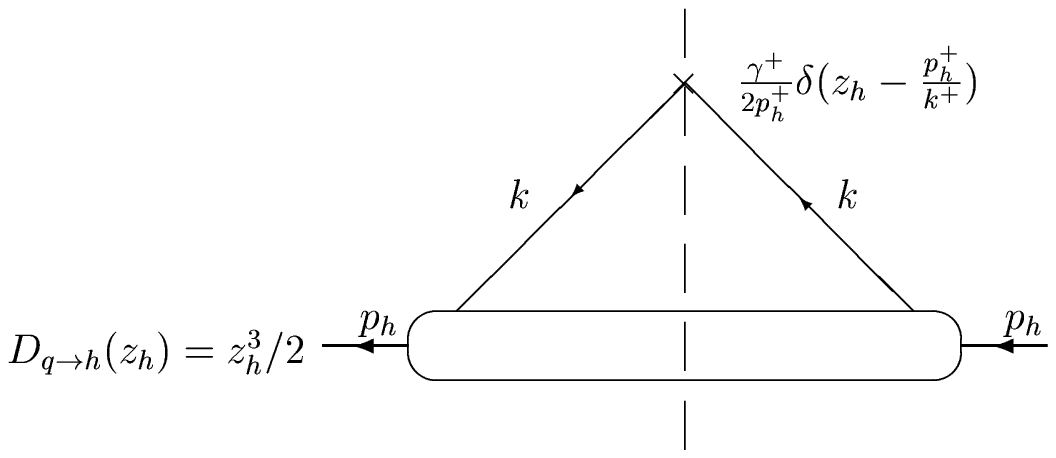,width=3.0in,height=1.5in}}
\centerline{(a)}
\centerline{\psfig{figure=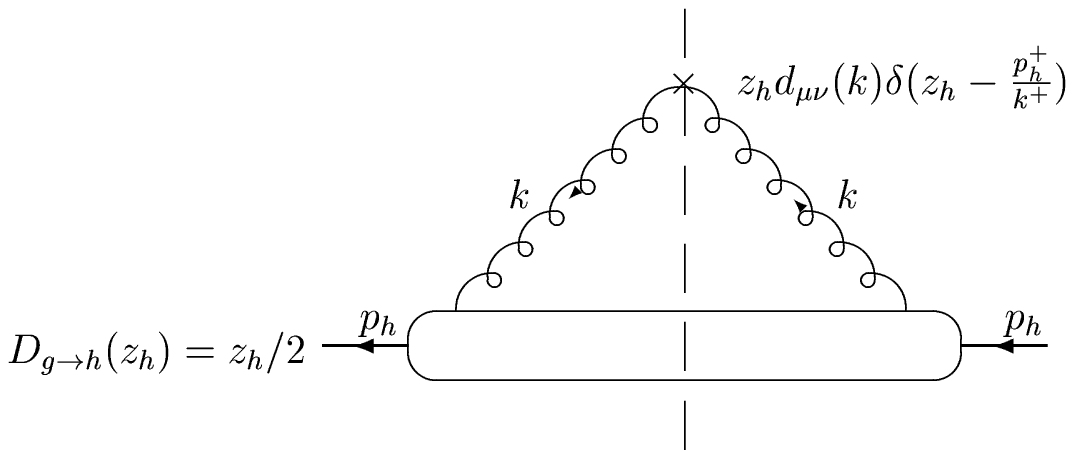,width=3.0in,height=1.5in}}
\centerline{(b)}
\caption{Cutting vertices for quark (a) and gluon (b) fragmentation function.}
\label{fig1}
\end{figure}

Using the `cut vertices' technique introduced by
A. Mueller\cite{Mueller78}, we can calculate
the scale dependence of these functions in perturbation theory.
The Feynman diagrams illustrating this calculation
are shown at leading order in Fig.~\ref{fig1}.  The corresponding
bare quark and gluon cut
vertices are
 \begin{mathletters}
 \label{cutvertex}
 \begin{equation}
 \frac{\gamma\cdot n}{2 p_h\cdot n}\delta\left(z_h {-}
     \frac{p_h\cdot n}{k\cdot n}\right)
 \label{cutvertex1}
\end{equation}
and
\begin{equation}
z_hd_{\mu\nu}(k)\delta\left(z_h-
      \frac{p_h\cdot n}{k\cdot n}\right)\;\; ,
 \label{cutvertex2}
 \end{equation}
 \end{mathletters}
\noindent respectively, in the light-cone gauge.

\begin{figure}
\centerline{\psfig{figure=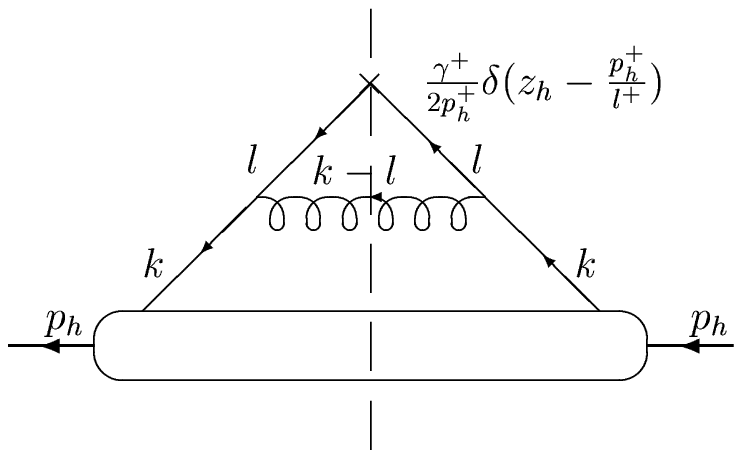,width=2.25in,height=1.5in}}
\caption{Real correction ($q\rightarrow q+g$) to the quark fragmentation.}
\label{fig2}
\end{figure}
\begin{figure}
\centerline{\psfig{figure=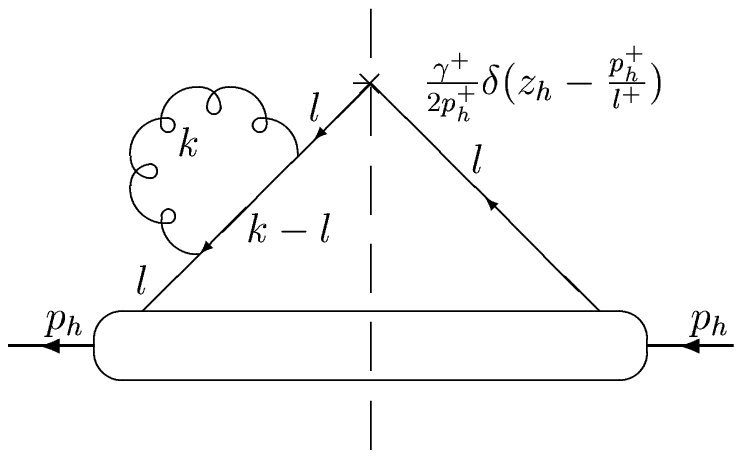,width=2.25in,height=1.5in}}
\caption{Virtual correction  to the quark fragmentation.}
\label{fig3}
\end{figure}

At next-to-leading order, ${\cal O}(\alpha_s)$, the gluon
bremsstrahlung process represented in Fig.~\ref{fig2}
gives a positive contribution 
to the quark fragmentation function.  This is due to the fact that
more options are open to the fragmenting quark at this 
order: it can fragment immediately, or radiate a 
gluon first.  This bremsstrahlung process becomes easier
as the energy of the radiated gluon decreases,
leading to a divergence as the energy goes to zero.
Physically, this divergence is meaningless.  If the 
quark radiates no energy, its fragmentation should not be
altered.  This idea is realized mathematically
by the self-energy contribution represented 
in Fig.~\ref{fig3}.  This correction removes the 
double-counting inherent in the above argument, and ensures that 
each decay channel is counted only once.
For this reason, the divergent contribution from zero-energy
bremsstrahlung is entirely canceled.

All next-to-leading order fragmentation 
sub-processes lead to similar corrections to the fragmentation
functions.  We will list a complete set of Feynman `cut' diagrams later.
Apart from the spurious infrared divergences
in some of the diagrams, each correction contains a collinear
divergence as the transverse momentum of the loop approaches zero.  
These divergences signal an unavoidable scale dependence 
in our corrections.  Physically, this scale dependence 
represents the fact that partons appear different when probed at different
scales.  We can handle these divergences mathematically
by computing only the difference between our fragmentation
functions probed at different scales.  Since the asymptotic contribution
to our loop integrals cannot depend on the external scales of the process,
the collinear divergences will always cancel in such differences.
With this in mind, we write the results
 \begin{mathletters}
 \label{renormfrag}
 \begin{eqnarray}
   D_{q\rightarrow h}(z_h, Q^2) &=& D_{q\rightarrow h}(z_h,\mu^2)\nonumber\\
   &&+\frac{\alpha_s(\mu^2)}{2\pi}\int_{\mu^2}^{Q^2}
\frac{dk_{\perp}^2}{k_{\perp}^2}
   \int_{z_h}^1\frac{dz}{z}
   \Bigl[\gamma_{qq}(z)
   D_{q\rightarrow h}\left(\frac{z_h}{z},\mu^2\right)
{+}\gamma_{qg}(z)
   D_{g\rightarrow h}\left(\frac{z_h}{z},\mu^2\right)\Bigr]\, ,
 \label{renormfrag1}\\
   D_{g\rightarrow h}(z_h, Q^2) &=& D_{g\rightarrow h}(z_h,\mu^2)
   \nonumber\\&&+\frac{\alpha_s(\mu^2)}{2\pi}
\int_{\mu^2}^{Q^2}\frac{dk_{\perp}^2}{k_{\perp}^2}
   \int_{z_h}^1\frac{dz}{z}
   \Bigl[\gamma_{gq}(z)
D_{s\rightarrow h}\left(\frac{z_h}{z},\mu^2\right)
{+}\gamma_{gg}(z)
   D_{g\rightarrow h}\left(\frac{z_h}{z},\mu^2\right)\Bigr]\, ,
 \label{renormfrag2}
 \end{eqnarray}
 \end{mathletters}
of the calculation.  Here, we have defined the 
singlet quark fragmentation function as,
\begin{equation}
D_{s\rightarrow h}(z,\mu^2)\equiv\sum_{q}\left(
D_{q\rightarrow h}(z,\mu^2)
+D_{\overline q\rightarrow h}(z,\mu^2)\right)\;\; .
\end{equation}
The integration kernels, or splitting functions, are
given by \cite{Field}:
 \begin{mathletters}
 \label{split}
 \begin{eqnarray}
  \gamma_{qq}(z)&=& C_F\Bigl[\frac{1+z^2}{(1-z)_+}
    +\frac{3}{2}\delta(1-z)\Bigr]\, ,
 \label{split1}\\
   \gamma_{qg}(z)&=& C_F\frac{1+(1-z)^2}{z}\, ,
 \label{split2}\\
  \gamma_{gq}(z)&=&T_F
    \Bigl[ z^2+(1-z)^2\Bigr]\, ,
 \label{split3}\\
   \gamma_{gg}(z)&=& 2C_A\Bigl[\frac{z}{(1-z)_+}
     +\frac{1-z}{z}+z(1-z)\Bigr]
     +\delta(1-z)\Bigl[\frac{11}{6}C_A-\frac{2}{3}n_f T_F\Bigr]\, .
 \label{split4}
 \end{eqnarray}
 \end{mathletters}
Here, $n_f$ is the number of active quark flavors and the 
SU($N_c$) Casimirs are given by
$C_F=(N_c^2-1)/2N_c$, $C_A=N_c$ and $T_F=1/2$; $N_c=3$. The `+'-function
is defined such that the replacement
 \begin{equation}
 \label{+function}
  \int_0^1 dz\frac{f(z)}{(1-z)_+}=\int_0^1 dz
   \frac{f(z)-f(1)}{1-z}\, 
 \end{equation}
is valid for any function $f(z)$ that is continuous at $z=1$.

The scale dependence of the
fragmentation functions is then determined by the 
DGLAP evolution equation \cite{dglap},
 \begin{mathletters}
 \label{DGLAP0}
 \begin{eqnarray}
   Q^2\,\frac{d}{d Q^2}D_{q\rightarrow h}(z_h, Q^2)
   &=&\frac{\alpha_s(Q^2)}{2\pi}\int_{z_h}^1\frac{dz}{z}
   \Bigl[\gamma_{qq}(z)
   D_{q\rightarrow h}(z_h/z, Q^2)
   +\gamma_{qg}(z)
   D_{g\rightarrow h}(z_h/z, Q^2)\Bigr]\, ,
 \label{DGLAP01}\\
   Q^2\,\frac{d }{d Q^2}D_{g\rightarrow h}(z_h, Q^2)
   &=&\frac{\alpha_s(Q^2)}{2\pi}\int_{z_h}^1\frac{dz}{z}
   \Bigl[\gamma_{gq}(z)
   D_{s\rightarrow h}(z_h/z, Q^2)
   +\gamma_{gg}(z)
   D_{g\rightarrow h}(z_h/z,Q^2)\Bigr]\, ,
 \label{DGLAP02}
 \end{eqnarray}
 \end{mathletters}
as can be shown by repeated iteration of two-particle-irreducible
diagrams similar to those shown in Figs. \ref{fig2} and \ref{fig3}.  
Eq.(\ref{renormfrag})
is analogous to the `Born' approximation of (\ref{DGLAP0}).
Since we have calculated our splitting functions only
at leading order in $\alpha_s$, our evolution equations 
are not correct to all orders.  However, 
it can be shown \cite{Field} that Eq.~(\ref{DGLAP0})
contains the highest power of $\log(Q^2/\mu^2)$
present at each order in $\alpha_s$.  For this 
reason, it is known as the leading log approximation (LLA)
to the full evolution equation.

\subsection{Fragmentation Function at Finite Temperature}
\label{sec2b}

\begin{figure}
\centerline{\psfig{figure=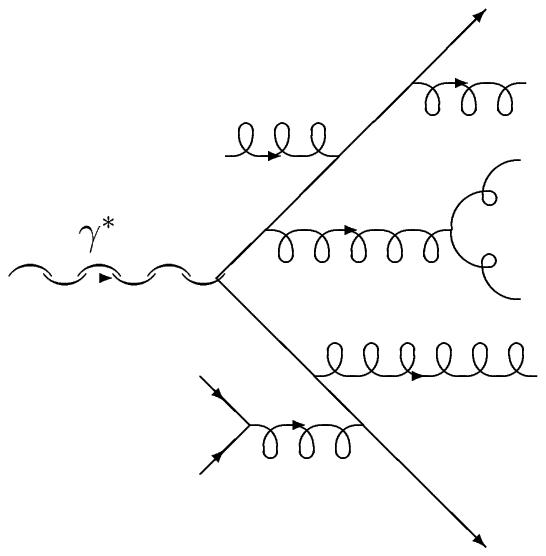,width=2.0in,height=2.0in}}
\caption{Parton emission and absorption by propagating jets in a medium.}
\label{fig4}
\end{figure}

In the finite-temperature case, we consider the fragmentation
of a parton in a thermal medium with temperature $T$.
In such a thermal environment, it will interact with the medium 
through emission and absorption as illustrated in Fig.~\ref{fig4}.
Assuming $N$ quarks and gluons are absorbed and $M$ quarks and 
gluons are emitted, we can express the hadronic tensor
[Eq.~(\ref{hadron1})] at finite temperature as
 \begin{eqnarray}
 \label{hadron2}
  W_{T\neq 0}^{\mu\nu}(q)&=\frac{1}{4\pi}&\int\frac{d^3p_1}{(2\pi)^3 2E_1}
    \cdots\frac{d^3p_N}{(2\pi)^3 2E_N}
    \frac{d^3p'_1}{(2\pi)^3 2E'_1}
    \cdots\frac{d^3p'_M}{(2\pi)^3 2E'_M}
    (2\pi)^4\delta(\sum_{i=1}^M p'_i-\sum_{i=1}^N p_i-q)
 \nonumber\\
  &&\times\langle p_1,\cdots, p_N|J^{\mu}(0)|p'_1\cdots p'_M\rangle
    \langle p'_1\cdots p'_M|J^{\nu}(0)|p_1,\cdots, p_N\rangle
 \nonumber\\
  & &\times\frac{1}{e^{\beta E_1}\mp 1}\cdots\frac{1}{e^{\beta E_N}\mp 1}
     \Bigl[1\pm\frac{1}{e^{\beta E'_1}\mp 1}\Bigr]\cdots
     \Bigl[1\pm\frac{1}{e^{\beta E'_M}\mp 1}\Bigr]\, .
 \end{eqnarray}
The phase space integrals of the
quarks and gluons are weighted by Bose-Einstein or Fermi-Dirac
thermal distributions for gluon and quark absorption, and by the
Bose-Einstein enhancement and Pauli blocking factor for emission.
As in the derivation of the
emission rate for photons and dileptons from a quark gluon plasma
by McLerran and Toimela\cite{MT85}, the hadron
tensor in Eq.~(\ref{hadron2}) can be expressed as the 
thermal expectation value of a
partonic electromagnetic current-current correlation function,
 \begin{equation}
 \label{hadron3}
  W_{T\neq 0}^{\mu\nu}(q)=\frac{1}{4\pi}\int d^4x e^{-i q\cdot x}
  \langle J^{\mu}(0)J^{\nu}(x)\rangle\, ,
 \end{equation}
where $\langle\cdots\rangle$ stands for the thermal expectation
value,
 \begin{equation}
 \label{expectation}
  \langle{\cal O}\rangle=\frac{{\rm Tr}[e^{-\beta {\hat H}}{\cal O}]}
    {{\rm Tr}\;e^{-\beta {\hat H}}}\, .
 \end{equation}

Comparing to the hadronic tensor of Eq.~(\ref{hadron1}) defined at zero
temperature, the hadronic tensor in Eq.~(\ref{hadron3}) at finite
temperature can be obtained by replacing the vacuum expectation
value in Eq.~(\ref{hadron1}) by a thermal expectation value.
Applying this idea to the fragmentation functions is slightly more
complicated.  
First, the conventional interpretation of the fragmentation functions
as probabilities for a parton to fragment into a hadron needs to be
modified. The thermal QCD medium is defined to be color deconfined,
so the concept of hadrons inside such a thermal medium is ill-defined.
Hadrons should only emerge after the hadronization of the medium. Therefore,
the fragmentation functions in a thermal medium are defined in such a way that
they give the probabilities of producing a hadron as the partons hadronize 
together with the medium. Another useful scenario is to consider how
an energetic parton with high virtuality produces secondary partons. In this
case, the fragmentation functions can be defined as the probabilities
of finding a secondary parton inside a parton jet. One can then study how
such parton production from a jet is modified in a thermal medium. 

At zero temperature, the absence of a scale 
implies that the fragmentation functions depend only
on the ratio of the hadron and parton energies.  
This leads one to consider the fraction of parton momentum 
carried by the hadron rather than the absolute momentum.
At finite temperature, the thermal average introduces
dependence on temperature and therefore the absolute 
momenta of the initial partons.  This implies that a 
fragmentation function defined in the spirit of the 
zero temperature functions cannot be complete.
In particular, fragmentation functions 
relating to different initial parton energies will
mix with each other under renormalization, 
resulting in a very complicated evolution equation.

A more straightforward way to approach the problem of 
fragmentation at finite temperature is to 
consider the momenta of the fragmenting parton
and the final state hadron separately.
In particular, 
\begin{equation}
\label{22}
F_{h/q}(y;x)={y\over4x}\;\int{d^{\,4}k\over(2\pi)^4}
\,\delta(k^+-x)\,\int d^{\,4}z e^{-ikz}
\sum_{\tilde S}
{\rm Tr}\left[ 
\langle\psi(0)|\tilde S,p_h\rangle\gamma^+\langle
\tilde S,p_h|\overline\psi(z)\rangle\right]
\end{equation}
represents the probability for a 
quark of momentum $p^+=x$ to decay into a 
hadron of momentum $p_h^+=y$.
This expression is to be understood as a thermal 
average over the quark field operators  
in which a particular intermediate state, $|p_h\rangle$,
is singled out.  The set of states $\{|\tilde S\rangle\}$ 
includes the thermal phase space indicated in Eq.~(\ref{hadron2})
and the quark color is implicitly averaged. Note that in the
following discussion of the newly defined fragmentation functions
at finite temperature, all variables, $x$, $y$, $z$, etc. represent
absolute '$+$'-momentum and thus carry the dimension of energy. 
We will denote fractional momentum by these variables with a 
subscript ($z_f$, for example).

This new function is quite similar in form to the 
quark fragmentation function in the vacuum. 
Defining
\begin{equation}
{\widetilde D}_{q\rightarrow h}(z_f;x)\equiv F_{h/q}(xz_f;x)\;\; ,
\label{Dqdef}
\end{equation}
we find that ${\widetilde D}_{q\rightarrow h}(z_f;x)|_{{T}=0}
=D_{q\rightarrow h}(z_f)$.  However, the interpretation
of $F_{h/q}$ is entirely different.  
By tracking the absolute
momentum of both the initial parton and the observed hadron,
this new function is able to clearly express the dependence
of fragmentation on the temperature and the initial parton energy. 
We will see below that this separation of fragmenting parton
and observed hadron is essential to a transparent
evolution equation at finite temperature where fragmentation
functions with different initial parton energies mix.

Fragmentation functions for anti-quarks and gluons can be
defined in a similar way.  We have
\begin{mathletters}
 \label{fragT}
 \begin{eqnarray}
F_{h/\overline q}(y;x)&=&{y\over4x}\;\int{d^{\,4}k\over(2\pi)^4}
\,\delta(k^+-x)\,\int d^{\,4}z e^{-ikz}
\sum_{\tilde S}
\langle\overline\psi(0)|\tilde S,p_h\rangle\gamma^+
\langle\tilde S,p_h|\psi(z)\rangle\label{fragT2}\\
F_{h/g}(y;x)&=&-{y\over2x^2}\;\int{d^{\,4}k\over(2\pi)^4}
\,\delta(k^+-x)\,\int d^{\,4}z e^{-ikz}
\sum_{\tilde S}
\langle G^{+\mu}(0)|\tilde S,p_h\rangle
\langle\tilde S,p_h|G^+_{\,\mu}(z)\rangle\;\; ,\label{fragT3}
\end{eqnarray}
\end{mathletters}
where colors are implicitly averaged over, as before.

\section{Radiative Corrections}
\label{sec3}

In this section we calculate the first order 
correction to the fragmentation
functions at finite temperature.  The result will be used
to derive the corresponding
evolution equations in the next section.  
Using
\begin{equation}
{y\over x}\delta(k^+-x)=z_h^3\,{1\over p_h\cdot n}\delta\left(
z_h-{p_h\cdot n\over k\cdot n}\right)\;\; ,
\end{equation}
where $z_h=y/x$ and $p_h\cdot n=y$, we see that the
form of this calculation is quite similar to 
that of the zero-temperature case.  
The only practical difference between the 
two cases lies in the form of the propagators,
both cut and uncut.  For uncut (intermediate state)
propagators, the expressions  
 \begin{mathletters}
 \label{propagator}
 \begin{eqnarray}
  S(k)&=&\Bigl[\frac{i}{k^2+i\eta}
   -2\pi f({k^+})
   \delta(k^2)\Bigr]{\not k}\, ,
 \label{propagator1}\\
  \Delta_{\mu\nu}(k)&=&\Bigl[\frac{i}{k^2+i\eta}
   +2\pi n({k^+})
   \delta(k^2)\Bigr]d_{\mu\nu}(k)\, ,
 \label{propagator2}
 \end{eqnarray}
 \end{mathletters}
replace the ordinary zero-temperature propagators.
The distribution functions
 \begin{mathletters}
 \label{distribution}
 \begin{eqnarray}
  f(x)&=&\frac{1}{e^{|x|/T}+1}\, ,
 \label{distribution1}\\
  n(x)&=&\frac{1}{e^{|x|/T}-1}\, ,
 \label{distribution2}
 \end{eqnarray}
 \end{mathletters}
represent the effect of the thermal medium on
free propagation.  

Cut (final state) propagators of thermal partons are 
represented by 
 \begin{mathletters}
 \label{cutpropagator}
 \begin{eqnarray}
  S^{cut}(k)&=&\Bigl[\theta(k^+)
  -f({k^+})\Bigr]{\not k}
  2\pi\delta(k^2)\, ,
 \label{cutpropagator1}\\
  \Delta^{cut}_{\mu\nu}(k)&=&\Bigl[\theta(k^+)
  +n({k^+})\Bigr]
  d_{\mu\nu}(k)2\pi\delta(k^2)\, ,
 \label{cutpropagator2}
 \end{eqnarray}
 \end{mathletters}
where $\theta(x)$ is the step function:
 \begin{equation}
 \label{theta}
  \theta(x)=\left\{
  \begin{array}{ll}
    0\, , &x<0 \\ 1\, , &x>0\, .
  \end{array}
  \right.
 \end{equation}
As before, 
the vertices and propagators on the right side of the
cut are the complex conjugates of the usual ones.

\subsection{First Order Correction to Quark Fragmentation Functions}
\label{sec3a}

\begin{figure}
\centerline{\psfig{figure=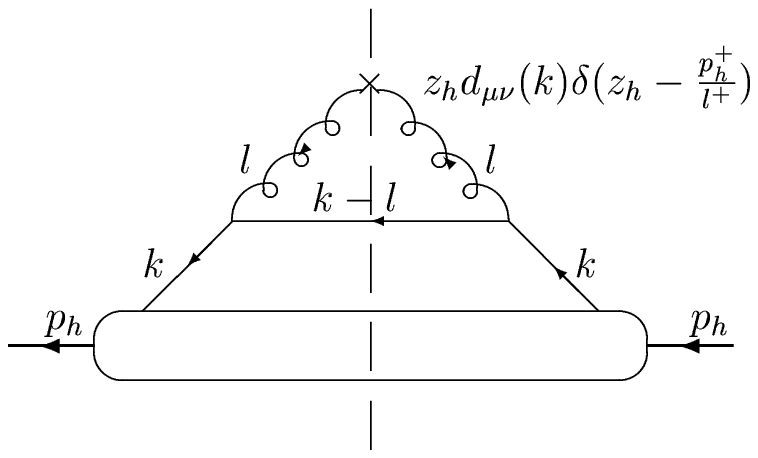,width=2.25in,height=1.5in}}
\caption{Real correction ($q\rightarrow g+q$) to the quark fragmentation.}
\label{fig5}
\end{figure}

The first order corrections to the quark fragmentation
function are illustrated in Figs.~\ref{fig2}, \ref{fig3} and \ref{fig5}. 
Figs.~\ref{fig2} and \ref{fig5} represent real gluon and quark 
emission and absorptive processes, which give a positive
contribution to the fragmentation function. Fig.~\ref{fig3} represents 
the virtual self-energy correction, which removes the double-counting 
inherent in the real corrections.  

The correction due to real gluon emission/absorption 
shown in Fig.~\ref{fig2} can be expressed as
 \begin{eqnarray}
 \label{dDqa1}
\Delta {F}_{h/q}^{(q)}(y;x)&=&
 {y^3\over4x^3}\,\int\,dz_f'\;
\frac{d^4 l}{(2\pi)^4}\delta\left(z_f'-{p_h^+\over l^+}\right)\nonumber\\
&&\times\int d^{\,4}\xi\,e^{-il\xi}
\sum_{\tilde S}{\rm Tr} \left[ \langle\psi(0)|\tilde S,p_h\rangle
\,{\gamma^+\over p_h^+}\langle\tilde S,p_h|\overline\psi(\xi)\rangle
\right]
{\rm Tr}\left\lbrack{\not\! p}_h\,
{\hat H}_q^{(q)}(l)\right\rbrack\, ,
 \end{eqnarray}
where the contribution from the loop is
 \begin{eqnarray}
 \label{hqa}
   {\hat H}_q^{(q)}(l)&=&C_F g^2
\int{d^{\,4}k\over(2\pi)^4} (-i\gamma_{\mu})
    \frac{i{\not k}}{k^2+i\eta}\left\lbrack\frac{\gamma^+}{2p_h^+}
\delta\left({y\over x}
-\frac{p_h^+}{k^+}\right)\right\rbrack
    \frac{-i{\not k}}{k^2-i\eta}(i\gamma_{\nu})
 \nonumber\\
   &&\times d^{\mu\nu}(k-l)\left[\theta(k^+ -l^+)
   +n({k^+ -l^+})\right] 2\pi\delta[(k-l)^2] \;\; .
 \end{eqnarray}
This expression describes a quark of momentum $k$
emitting a gluon of momentum $k-l$ (or absorbing a gluon of momentum
$l-k$), then
fragmenting into a hadron of momentum $p_h$ such that
$p_h^+/k^+=y/x$. Note that the thermal part of the uncut quark propagators
on each side of the cut does not contribute because the quark lines
are off-shell ($k^2\neq 0$) before the gluon radiation. This will be
the case for all other diagrams including the virtual corrections
after carefully taking the collinear approximation limit.

As it is, this expression cannot be written in terms of 
the simple collinear fragmentation function $F_{h/q}$
defined above.  The integral over $l$ couples its 
perturbative and nonperturbative parts 
in a way that $F_{h/q}$ is not sufficient to describe.
However, it can be shown \cite{factproof}
that the contributions to this expression from 
$l_\perp$ and $l^-$ are suppressed by powers of 
$p_h^+$ as $p_h^+\rightarrow\infty$.  At high energy,
we can expand $\hat H(l)$ about $l=[l^+,0,0_\perp]$ 
and drop the power-suppressed corrections.  This collinear
approximation captures the complete leading-twist 
behavior of our fragmentation, so it can be used to extract the 
evolution of the finite-temperature fragmentation
functions at the leading-twist level.

Using the collinear approximation,
 \begin{equation}
 \label{collinear}
   l=l_c\equiv[y/z_f',0,0_\perp]\, ,
 \end{equation}
in Eq.~(\ref{hqa}) decouples the 
integral over $l$ in Eq.~(\ref{dDqa1}) from
$\hat H(l_c)$.  This allows us to write the correction as
\begin{equation}
\Delta {F}_{h/q}^{(q)}(y;x)=
 \int\,dz_f'\;\left({y\over xz_f'}\right)^3\,F_{h/q}(y;y/z_f')
{\rm Tr}\left\lbrack{\not\! p}_h\,
{\hat H}_q^{(q)}(l_c)\right\rbrack\;\; .
\end{equation}
Performing the change
\begin{equation}
 \label{z}
   z_f'={y\over z}
 \end{equation} 
in $\hat H(l_c)$,
we arrive at the result
 \begin{equation}
 \label{dDqa2}
\Delta {F}_{h/q}^{(q)}(y;x)
 =\frac{\alpha_s}{2\pi}\int\frac{dk_{\perp}^2}{k_{\perp}^2}
   \int_0^{\infty}\frac{dz}{z}
   P_{qq}(z/x)\epsilon(x-z)
\lbrack \theta(x-z)+n(x-z)
\rbrack F_{h/q}(y;z)\;\; .
 \end{equation}
Here, we have defined
 \begin{equation}
 \label{sign}
  \epsilon(x)=\left\{
  \begin{array}{ll}
    -1\, , &x<0 \\ +1\, , &x>0\, 
  \end{array}
  \right.
 \end{equation}
and 
 \begin{equation}
 \label{Pqqg}
   P_{qq}(z_f)=C_F\frac{1+z^2_f}{1-z_f}\, .
 \end{equation}

The integral over $z$ in Eq.(\ref{dDqa2}) can be divided into two
regions: $0<z<x$ and $z>x$.  In the first region,
we have the physical situation in which the quark 
emits a gluon into the thermal bath before it 
fragments.  The second region represents a physical
absorption process.  In either case, 
the plus-component of the intermediate quark
momentum is $z$.
At zero temperature, 
only the emission process contributes.  
Since the quark can only lose energy through
evolution in this case, the `+'-momentum 
of the intermediate quark must be larger
than that of the observed hadron as well as 
smaller than that of the initial quark, $y<z<x$.
This leads to the limits on the zero-temperature
evolution equation in Eq.~(\ref{DGLAP0}).  At finite 
temperature, the quark can gain energy
through absorption.  This process 
extends the region of integration in both 
directions, $z>x$ and $z<y$.
The extension to $z>x$ is an obvious effect of 
absorption: the intermediate quark 
in an absorptive process has more energy
than the initial quark.  The 
extension $z<y$ is somewhat more subtle.  At
zero temperature, conservation of energy implies that the 
observed hadron cannot have more energy than its parent
parton.  At finite temperature, we have no such 
restriction.  The thermal bath can provide the 
excess energy needed for a parton to fragment into
a more energetic hadron.  This can be 
thought of as a kind of `collective' fragmentation
of the parton and the bath.  We can immediately
see from Eq.~(\ref{dDqa2}) that the inclusion 
of this effect is necessary: even if we choose
to take $F_{h/q}(y;x)=0$ for $y>x$,
Eq.~(\ref{dDqa2}) will generate a finite contribution in this region.

Our phenomenological interpretation 
of Eq.~(\ref{dDqa2}) leads to a somewhat 
fundamental difficulty with the definition 
of fragmentation functions at finite temperature.
As the intermediate energy, $z$, 
decreases around and below the temperature, 
it becomes inconsistent to treat the parton
as separate from the bath itself.  Furthermore,
it is impossible to distinguish the fragmentation
of a low-energy parton in a thermal bath 
from the fragmentation of the bath itself.
This difficulty can be resolved via a
phenomenological separation between the 
fragmentation of a parton distinct from the bath 
and that of the bath itself,
as we will discuss in the next section.  For the remainder of this
section, we will ignore our difficulty and take the 
radiative corrections at face value.

As in the zero-temperature case, our result
for real emission/absorption is plagued with 
infrared divergences as $z\rightarrow x$.
The divergences in Eq.~(\ref{dDqa2}) are made
worse by the Bose-Einstein functions.  As $z\rightarrow x$,
these functions generate a linear divergence in addition
to the logarithmic divergence present in the 
zero-temperature part.  All of these divergences 
are unphysical in the sense that they represent 
overcounting of the relevant 
degrees of freedom.  As explained above, they must
be canceled by virtual corrections.

The virtual correction shown in Fig.~\ref{fig3} reads
 \begin{eqnarray}
 \label{dDqb1}
\Delta {F}_{h/q}^{(q)}(y;x)&=&
 {y^3\over4x^3}\,\int\,dz_f'\;
\frac{d^4 l}{(2\pi)^4}\delta\left(z_f'-{p_h^+\over l^+}\right)\nonumber\\
&&\times\int d^{\,4}\xi\,e^{-il\xi}
\sum_{\tilde S}{\rm Tr} \left[\langle\psi(0)|\tilde S,p_h\rangle
\,{\gamma^+\over p_h^+}
\langle\tilde S,p_h|\overline\psi(\xi)\rangle\right]
{\rm Tr}\left\lbrack{\not\! p}_h\,
{\hat H}_q^{(v)}(l)\right\rbrack\, ,
 \end{eqnarray}
where 
 \begin{eqnarray}
 \label{hqb}
   {\hat H}_q^{(v)}(l)&{=}&\frac{1}{2}
   \Bigl[\Sigma(l)\frac{i{\not l}}{l^2+i\eta}{+}
   \Sigma^*(l)\frac{-i{\not l}}{l^2-i\eta}\Bigr]
   \frac{\gamma^+}{2p_h^+}\delta\left({y\over x}
{-}\frac{p_h^+}{l^+}\right)
 \nonumber\\
   &{=}&-{\rm Im}\Bigl[\frac{\Sigma(l)}{l^2+i\eta}\Bigr]{\not l}
   \frac{\gamma^+}{2p_h^+}\delta\left({y\over x}
{-}\frac{p_h^+}{l^+}\right)\;\; .
 \end{eqnarray}
Here, $\Sigma(l)$ is the quark self-energy.
The prefactor of $1/2$ comes from the renormalization of the
initial state \cite{CQ89}.

As before, we intend to use the collinear approximation
in Eq.~(\ref{collinear}) to relate the nonperturbative
part of this expression to $F_{h/q}$.  In this
case, however, direct substitution of $l=l_c$
leads to an indeterminate form, since $\Sigma(l)/l^2$ is not 
defined for $l^2=l_c^2$.  For this reason, 
we must carefully take the limit of Eq.~(\ref{hqb})
as $l\rightarrow l_c$.  With this in mind, 
we can read the quark self-energy at one-loop 
from the diagram in Fig.~\ref{fig3}: 
 \begin{equation}
 \label{qselfe1}
    \Sigma(l)=-g^2C_F\int\frac{d^4 k}{(2\pi)^4}
    \gamma_{\rho} S(l-k)\gamma_{\sigma}\Delta^{\rho\sigma}(k)\, .
 \end{equation}
After completing the $k^-$ integral by contour, the collinear divergent 
part of this expression is given by
 \begin{eqnarray}
 \label{qselfe2}
  &&\lim_{l\rightarrow l_c}{1\over l^2}\,{\rm Tr}\Bigl[{\not l}\Sigma(l)
{\not l}{\not n}\Bigr]
   =2i\,\frac{\alpha_s}{\pi} l_c^+\int\frac{dk_{\perp}^2}{k_{\perp}^2}
     \left\lbrace\int_{-\infty}^{\infty} {dz\over x} P_{qq}(z/x)
   \left\lbrack 
\theta(z)\theta(x-z)+n(x-z)
   \epsilon(x-z)-f(z)
   \epsilon(z)\right\rbrack\right\rbrace
 \nonumber\\
  &&\quad=2i\frac{\alpha_s}{\pi} l_c^+
\int\frac{dk_{\perp}^2}{k_{\perp}^2}
   \left\lbrace\int_0^x {dz\over x} P_{qq}(z/x)
   \left\lbrack\left(1+n(x-z)\right)
   \left(1-f(z)\right)
   +n(x-z)
   f(z)\right\rbrack\right.
 \nonumber\\
   &&\qquad\qquad\qquad
   +\int_{-\infty}^0 {dz\over x} P_{qq}(z/x)
   \left\lbrack\left(1+n(x-z)
   \right)f(z)
   +n(x-z)
   \left(1-f(z)\right)\right\rbrack
 \nonumber\\
   &&\qquad\qquad\qquad\left.
   +\int_x^{\infty} {dz\over x}\; {z\over x}
 P_{qq}(x/z)
   \left\lbrack n(z-x)
   \left(1-f(z)\right)
   +\left(1+n(z-x)\right)
   f(z)\right\rbrack\right\rbrace\;\; ,
 \end{eqnarray}
where we have taken $l=l_c$
wherever it does not generate a singularity
and used the property 
$P_{qq}(z_f)=-z_fP_{qq}(1/z_f)$ of the quark splitting
function.  This result can be understood in terms of
physical decay and scattering processes at
finite temperature\cite{Weldon83}. In the second equality above,
the first line corresponds to the decay of a quark into a quark
and a gluon, as well as the inverse process.  The second and third
lines correspond to the forward and backward processes of Landau
damping via scattering of the quark with another thermal quark or
gluon from the thermal bath. Landau damping requires the
presence of thermally excited states, and disappears at zero
temperature. In each term, every incoming gluon (quark) from the
thermal bath is weighted with a Bose-Einstein distribution function
$n(k^+_i/T)$ [a Fermi-Dirac distribution function
$f(k^+_i/T)$], and each outgoing gluon (quark) receives a
Bose-Einstein enhancement factor $1+n(k^+_i/T)$ [a Pauli
blocking factor $1-f(k^+_i/T)$].

Substituting Eqs.(\ref{hqb}) and (\ref{qselfe2}) into
(\ref{dDqb1}) and collecting terms, we arrive at the virtual correction
 \begin{equation}
 \label{dDqb2}
  \Delta {F}_{h/q}^{(v)}(y;x)
  =-\frac{\alpha_s}{2\pi}{F}_{h/q}(y;x)
   \int\frac{dk_{\perp}^2}{k_{\perp}^2}
   \int_{-\infty}^{\infty} {dz\over x}
   \left\lbrack \theta(z)\theta(x-z)+
\epsilon(x-z)n(x-z)
   -\epsilon(z)f(z)\right\rbrack
   P_{qq}(z/x)\, .
 \end{equation}
In order to see clearly that all infrared divergences in the real
gluon radiation are canceled by the virtual correction, we 
combine Eqs.(\ref{dDqa2}) and (\ref{dDqb2}) and divide the
integral into different regions:
 \begin{eqnarray}
 \label{dDqab}
 \Delta {F}_{h/q}^{(q)}(y;x)
  +\Delta {F}_{h/q}^{(v)}(y;x)
  &=&\frac{\alpha_s}{2\pi}\int\frac{dk_{\perp}^2}{k_{\perp}^2}
   \left\lbrace\int_0^x\frac{dz}{z}
   P_{q\rightarrow qg}(z/x)
   \underbrace{\lbrack 1
   +n(x-z)\rbrack]}_{(a)}
{F}_{h/q}(y;z)\right.
 \nonumber\\
   &&+\int_x^{\infty} {dz\over x}
   P_{qq}(x/z)
   \underbrace{n(z-x)}_{(b)}
   {F}_{h/q}(y;z)
 \nonumber\\
  &&-{F}_{h/q}(y;x)\left\lbrack
   \int_0^x {dz\over x} P_{qq}(z/x)
   \lbrack\underbrace{1+n(x-z)}_{(c)}
   -\underbrace{f(z)}_{(d)}\rbrack
\right. \nonumber\\
   &&+\int_{-\infty}^0 {dz\over x} P_{qq}(z/x)
   \underbrace{\lbrack n(x-z)
   +f(z)\rbrack}_{(e)}
 \nonumber\\
  &&\left.\left.+\int_x^{\infty} {dz\over x}\, {z\over x}
   P_{qq}(x/z)
   \lbrack\underbrace{n(z-x)}_{(f)}
   +\underbrace{f(z)}_{(g)}\rbrack
   \right\rbrack\right\rbrace\;\; . \label{eq:cancel}
 \end{eqnarray}
As $z\rightarrow x$, the logarithmic divergences in
terms $(a)$, $(b)$ and $(d)$ are canceled by those in 
$(c)$, $(f)$ and $(g)$, respectively.  
The linear infrared
divergences in terms $(a)$ and $(b)$ 
are canceled by those in $(c)$ and $(f)$, 
leaving behind a residual
logarithmic divergence.  This divergence 
cancels in the sum of terms $(a)$ and $(b)$
as long as $F_{h/q}(y;z)$ is differentiable at $z=x$.
Term (e) is finite.

Another real correction to quark fragmentation comes from Fig.~\ref{fig5},

 \begin{eqnarray}
\Delta {F}_{h/q}^{(g)}(y;x)&=&
 {y \over4x^3}\,\int\,dz_f'\;
\frac{d^4 l}{(2\pi)^4}\delta\left(z_f'-{p_h^+\over l^+}\right)(-{z_f'}^2) 
\nonumber\\
&&\times\int d^{\,4}\xi\,e^{-il\xi}
\sum_{\tilde S} \langle G^{+\mu}(0)|\tilde S,p_h\rangle
\langle\tilde S,p_h|G^+_{\mu}(\xi)\rangle\,\,
d_{\mu\nu}(l){\hat H}_q^{\mu\nu(g)}(l)\, ,
 \end{eqnarray}
Here, the loop gives the hard contribution
 \begin{eqnarray}
 \label{hqc}
{\hat H}_q^{\mu\nu(g)}(l_c)&=& C_F g^2
\int{d^{\,4}k\over(2\pi)^4}
    {\rm Tr}\Bigl[ (-i\gamma^{\mu})
    \frac{i{\not k}}{k^2+i\eta}\frac{\gamma^+}{2p_h^+}
    \frac{-i{\not k}}{k^2-i\eta}(i\gamma^{\nu})
    ({\not k}-{\not l_c})\Bigr]
 \nonumber\\
   &&\quad\times\left[\theta(k^+ {-}l_c^+)
   {-} f(k^+ {-}l_c^+)\right]
2\pi\delta\left[k^2-2l_c\cdot k\right]\delta\left({y\over x}{-}
\frac{p_h^+}{k^+}\right)\;\; .
 \end{eqnarray}
Performing the integrals and simplifying, we obtain
 \begin{equation}
 \label{dDqc2}
\Delta {F}_{h/q}^{(g)}(y;x)
  =\frac{\alpha_s}{2\pi}\int\frac{dk_{\perp}^2}{k_{\perp}^2}
   \int_0^{\infty}\frac{dz}{z}
   P_{qg}(z/x)\epsilon(x-z)
   \left\lbrack\theta(x-z)-f(x-z)\right\rbrack
   {F}_{h/g}(y;z)\;\; .
 \end{equation}
The new splitting function,
 \begin{equation}
 \label{Pqgq}
   P_{qg}(z_f)=\gamma_{qg}(z_f)
   =C_F\frac{1+(1-z_f)^2}{z_f}\;\; ,
 \end{equation}
is related to $P_{qq}(z_f)$ via $z_f\rightarrow 1-z_f$.

As in the zero temperature case, the scale dependence of
the fragmentation functions is generated by the collinear 
divergences present in Eqs.~(\ref{dDqab}) and (\ref{dDqc2}).  
To remove the divergences, we consider the 
difference between fragmentation functions
measured at two different scales:
 \begin{eqnarray}
 \label{dDq}
 {F}_{h/q}(y;x, Q^2)-
  {F}_{h/q}(y;x,\mu^2)&=&
  \Delta {F}_{h/q}^{(g)}(y;x, Q^2)+
  \Delta {F}_{h/q}^{(q)}(y;x, Q^2)+
  \Delta {F}_{h/q}^{(v)}(y;x, Q^2)
 \nonumber\\
 &=&
 \frac{\alpha_s(\mu^2)}{2\pi}
\int_{\mu^2}^{Q^2}\frac{dk_{\perp}^2}{k_{\perp}^2}
   \int_0^{\infty}\frac{dz}{z}
   \left\lbrack{\widetilde \gamma}_{qq}(z;x)
   {F}_{h/q}(y;z,\mu^2)+
   {\widetilde \gamma}_{qg}(z;x)
   {F}_{h/g}(y;z,\mu^2)\right\rbrack\;\; .
 \end{eqnarray}
The modified kernels are
 \begin{mathletters}
 \label{modisplita}
 \begin{eqnarray}
 {\widetilde \gamma}_{qq}(z;x)&=&P_{qq}(z/x)\epsilon(x-z)\lbrack
\theta(x-z)+n(x-z)\rbrack\nonumber\\
&&\qquad-\delta(1-x/z)
\int_{-\infty}^\infty{d\omega\over x}\,P_{qq}(\omega/x)
\left\lbrack \theta(\omega)\theta(x-\omega)+\epsilon(x-\omega)n(x-\omega)
-\epsilon(\omega)f(\omega)\right\rbrack
 \label{modisplita1}\\
{\widetilde \gamma}_{qg}(z;x)&=&
P_{qg}(z/x)\epsilon(x-z)\left\lbrack
\theta(x-z)-f(x-z)\right\rbrack\;\; ,  
 \label{modisplita2}
 \end{eqnarray}
 \end{mathletters}
where the splitting functions are given by 
Eqs.(\ref{Pqqg}) and (\ref{Pqgq}).  
Obviously, the zero-temperature result 
is recovered as $n$ and $f$ approach zero.

\subsection{First Order Correction to Gluon Fragmentation Functions}
\label{sec3b}

\begin{figure}
\centerline{\psfig{figure=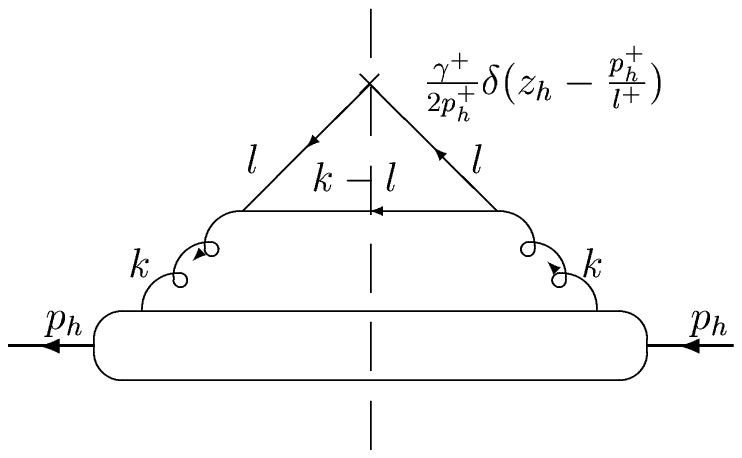,width=2.25in,height=1.5in}}
\caption{Real correction ($g\rightarrow q+\bar{q}$) to the gluon 
fragmentation.}
\label{fig6}
\end{figure}

\begin{figure}
\centerline{\psfig{figure=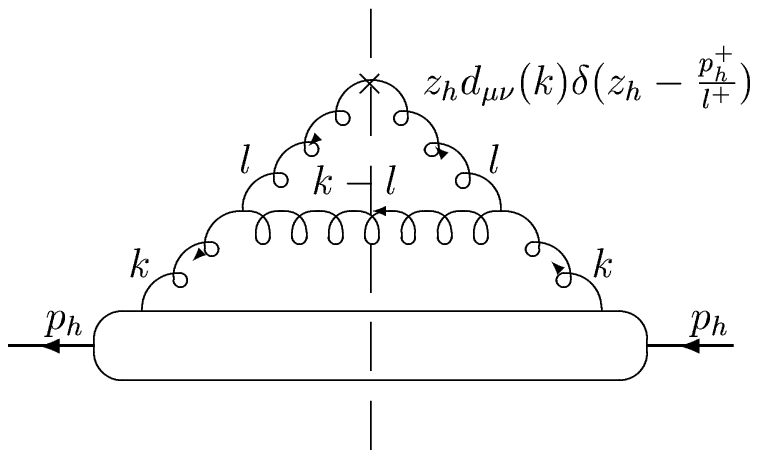,width=2.25in,height=1.5in}}
\caption{Real correction ($g\rightarrow g+g$) to the gluon fragmentation.}
\label{fig7}
\end{figure}

\begin{figure}
\centerline{\psfig{figure=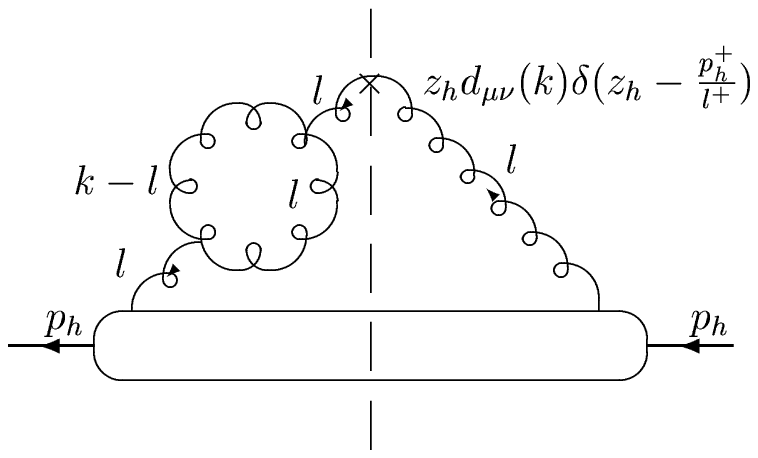,width=2.25in,height=1.5in}}
\caption{Virtual correction (gluon-loop) to the gluon fragmentation.}
\label{fig8}

\end{figure}

\begin{figure}
\centerline{\psfig{figure=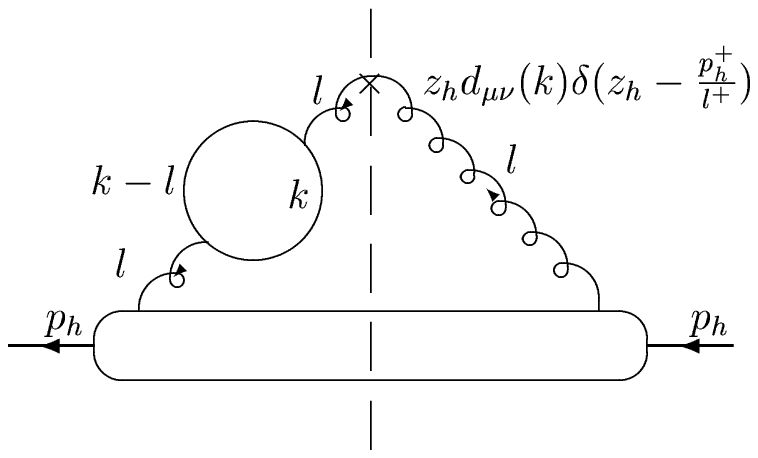,width=2.25in,height=1.5in}}
\caption{Virtual correction (quark-loop) to the gluon fragmentation.}
\label{fig9}
\end{figure}

In the singlet sector, the calculation of first order correction
to the gluon fragmentation function is quite similar.  
The first order real and
virtual corrections to the gluon fragmentation function are 
illustrated by Feynman `cut' 
diagrams in Figs.~\ref{fig6}-\ref{fig9}.



The real correction from Fig.~\ref{fig6} with gluon splitting into
quark and anti-quark pair is,

\begin{eqnarray}
\Delta {F}_{h/g}^{(q)}(y;x)&=&
 {y\over4x}\,\int\,dz_f'\;
\frac{d^4 l}{(2\pi)^4}\delta\left(z_f'-{p_h^+\over l^+}\right)
\int d^{\,4}\xi\,e^{-il\xi} \sum_{q=1}^{n_f} \sum_{\tilde S}
\left\{ {\rm Tr} \left[ \langle\psi(0)|\tilde S,p_h\rangle
\,{\gamma^+\over p_h^+}\langle\tilde S,p_h|\overline\psi(\xi)\rangle
\right] \right.
\nonumber\\
&+&
\left.\langle\overline\psi(0)|\tilde S,p_h\rangle
\,{\gamma^+\over p_h^+}\langle\tilde S,p_h|\psi(\xi)\rangle\right\}
{\rm Tr}\left\lbrack{\not\! p}_h\,
{\hat H}_g^{(q)}(l)\right\rbrack\, ,
 \end{eqnarray}

which contains the hard loop
correction
 \begin{eqnarray}
 \label{hga}
 {\hat H}_g^{(q)}(l_c)&=&T_F g^2 \int{d^{\,4}k\over(2\pi)^4}
(-i\gamma_{\mu})
    ({\not k}-{\not l}_c)(i\gamma_{\nu})
 \nonumber\\
  &&\quad\times\left[\theta(k^+ -l_c^+)
    -f\left(k^+ -l_c^+\right)\right]
    2\pi\delta\left[k^2-2l_c\cdot k\right]
 \nonumber\\
  &&\quad\times\left(\frac{id^{\mu\alpha}(k)}{k^2+i\eta}\right)
    \left(\frac{-id^{\nu\beta}(k)}{k^2-i\eta}\right)
    {y\over x}\,d_{\alpha\beta}(k)
    \delta\left({y\over x}-\frac{p_h^+}{k^+}\right)\, .
 \end{eqnarray}
Using
 \begin{equation}
 \label{identity1}
   d^{\mu\alpha}(k)d_{\alpha\beta}(k)d^{\beta\nu}(k)
   =d^{\mu\nu}(k)\, .
 \end{equation}
to simplify Eq.~(\ref{hga})
and performing the $k^+$ and $k^-$ integrals
leaves us with the correction
 \begin{eqnarray}
 \label{dDga2}
 \Delta {F}_{h/g}^{(q)}(y;x)
   &=&\frac{\alpha_s}{2\pi}\int\frac{dk_{\perp}^2}{k_{\perp}^2}
   \int_0^{\infty}\frac{dz}{z}
   P_{gq}(z/x)\epsilon(x-z)
   \lbrack\theta(x-z)-f(x-z)\rbrack
   F_{h/s}(y;z)
 \end{eqnarray}
to $F_{h/g}(y;x)$.  The splitting function
 \begin{equation}
 \label{Pgqq}
   P_{gq}(z_f)=\gamma_{gq}(z_f)=
   T_F\left[z_f^2+(1-z_f)^2\right]\, 
 \end{equation}
is the same as the zero-temperature result.
Here, we have defined the singlet fragmentation
function
\begin{equation}
F_{h/s}(y;x)\equiv\sum_{q=1}^{n_f}\left({F}_{h/q}(y;x)
+F_{h/\overline q}(y;x)\right)\;\; .
\end{equation}
This expression is devoid of infrared divergences.

The contribution from Fig.~\ref{fig7} is
\begin{eqnarray}
\Delta {F}_{h/g}^{(g)}(y;x)&=&
 {1 \over4xy}\,\int\,dz_f'\;
\frac{d^4 l}{(2\pi)^4}\delta\left(z_f'-{p_h^+\over l^+}\right)(-{z_f'}^2) 
\nonumber\\
&&\times\int d^{\,4}\xi\,e^{-il\xi}
\sum_{\tilde S} \langle G^{+\mu}(0)|\tilde S,p_h\rangle
\langle\tilde S,p_h|G^+_{\mu}(\xi)\rangle\,\,
d_{\alpha\beta}(l){\hat H}_g^{\alpha\beta(g)}(l)\, ,
 \end{eqnarray}
where the hard sub-process represented is
 \begin{eqnarray}
 \label{hgb}
   {\hat H}_g^{\alpha\beta(g)}(l_c)
    &=&-C_A g^2 
\int{d^{\,4}k\over(2\pi)^4}V^{\beta\sigma\mu}(l_c, k-l_c, -k)
     V^{\alpha\lambda\nu}(-l_c, l_c-k, k)
 \nonumber\\
  &&\times\left[\theta(k^+ -l_c^+)
    +n\left(k^+ -l_c^+\right)\right]
    d_{\lambda\sigma}(k-l_c)
 \nonumber\\
  &&\times 2\pi\delta[k^2-2l_c\cdot k]
    \left(\frac{-id_{\mu\rho}(k)}{k^2-i\eta}\right)
    \left(\frac{id_{\tau\nu}(k)}{k^2+i\eta}\right)
 \nonumber\\
   &&\times {y\over x}d^{\rho\tau}(k)
    \delta\left({y\over x}-\frac{p_h^+}{k^+}\right)\, .
 \end{eqnarray}
The three-point vertex, $V^{\alpha\beta\gamma}(p,q,r)$,
is given by
 \begin{equation}
\label{V}
V^{\alpha\beta\gamma}(p,q,r)=
g^{\alpha\beta}(p-q)^\gamma+g^{\beta\gamma}(q-r)^\alpha
+g^{\gamma\alpha}(r-p)^\beta\;\; .
 \end{equation}

In order to perform the contraction of the Lorentz indices in
Eq.(\ref{hgb}), it is convenient to use Eq.(\ref{identity1}) 
along with the identities
 \begin{mathletters}
 \label{identity2}
 \begin{eqnarray}
   &&d_{\alpha\beta}(p) p^{\alpha}
    =d_{\alpha\beta}(p) p^{\beta}=0\, ,
 \label{identity2a}\\
   &&d_{\alpha\beta}(p) d^{\alpha\beta}(q)
    =2\, .
 \label{identity2b}
 \end{eqnarray}
 \end{mathletters}
Simplifying, we obtain
 \begin{eqnarray}
 \label{contraction1}
   &&d_{\alpha\beta}(l) V^{\beta\sigma\mu}(l, k-l, -k)
    V^{\alpha\lambda\nu}(-l, l-k, k)
    d_{\lambda\sigma}(k-l)
    d_{\mu\rho}(k)d^{\rho\tau}(k)d_{\tau\nu}(k)
 \nonumber\\
   &&\quad= d_{\alpha\beta}(l) V^{\beta\sigma\mu}(l, k-l, -k)
    V^{\alpha\lambda\nu}(-l, l-k, k)
    d_{\lambda\sigma}(k-l)
    d_{\mu\nu}(k)
 \nonumber\\
    &&\quad= -8k^2\left[\frac{z_f}{1-z_f}+\frac{1-z_f}{z_f}
    +z_f(1-z_f)\right]\, ,
 \end{eqnarray}
\noindent where $z_f=y/xz_f'$ as before. Performing the $k^+$
and $k^-$ integrations leaves us with
 \begin{eqnarray}
 \label{dDgb2}
 \Delta {F}_{h/g}^{(g)}(y;z)
   &=&\frac{\alpha_s}{2\pi}\int\frac{dk_{\perp}^2}{k_{\perp}^2}
    \int_0^{\infty}\frac{dz}{z}
    P_{gg}(z/x)\epsilon(x-z)
   \lbrack\theta(x-z)+n(x-z)\rbrack
F_{h/g}(y;z)\, .
 \end{eqnarray}

The splitting function
 \begin{equation}
 \label{Pggg}
   P_{gg}(z_f)
   =2C_A\left[\frac{z_f}{1-z_f}+\frac{1-z_f}{z_f}
    +z_f(1-z_f)\right]\, 
 \end{equation}
is familiar from the zero-temperature result.

As $z\rightarrow x$, our correction in Eq.~(\ref{dDgb2})
generates infrared divergences which must be 
canceled by virtual contributions.  
The loop in the diagram of Fig.~\ref{fig8} 
represents
 \begin{eqnarray}
 \label{hgc}
 {\hat H}_g^{\mu\sigma(vg)}(l)&=&\frac{1}{2}
   \Bigl[\Pi_{g}^{\mu\nu}(l)\frac{i}{l^2+i\eta}{+}
   \Pi_{g}^{\mu\nu*}(l)\frac{-i}{l^2-i\eta}\Bigr]
{y\over x}d_{\nu\rho}(l)d^{\rho\sigma}(l)
   \delta\left({y\over x}{-}\frac{p_h^+}{l^+}\right)
 \nonumber\\
   &=&-{\rm Im}\Bigl[\frac{\Pi_{g}^{\mu\nu}(l)}{l^2+i\eta}\Bigr]
   {y\over x} d_{\nu\rho}(l)d^{\rho\sigma}(l)
   \delta\left({y\over x}{-}\frac{p_h^+}{l^+}\right)\, .
 \end{eqnarray}
$\Pi_g^{\mu\nu}(l)$ is the non-Abelian contribution
to the gluon self-energy, which can be
written as
 \begin{equation}
 \label{gselfe1}
    \Pi_g^{\mu\nu}(l)=\frac{1}{2}g^2C_A\int\frac{d^4 k}{(2\pi)^4}
    V^{\beta\sigma\mu}(k, l-k,-l)
   V^{\alpha\lambda\nu}(-k, k-l, l)
    \Delta_{\alpha\beta}(k)\Delta_{\lambda\sigma}(l-k)\;\; .
 \end{equation}
The symmetry factor $1/2$ takes into
account the indistinguishability of the intermediate
gluons.  As with the quark virtual correction, we must
be careful when taking the limit $l\rightarrow l_c$.  
The result of the calculation contains the collinear
divergence
 \begin{eqnarray}
 \label{gselfe2}
  \lim_{l\rightarrow l_c}
{1\over l^2}\;d_{\sigma\mu}(l)\left\lbrack\Pi_{g}^{\mu\nu}(l)
\right\rbrack
    d_{\nu\rho}(l)d^{\rho\sigma}(l)
    &=& \lim_{l\rightarrow l_c}{1\over l^2}\;
\left\lbrack\Pi^{\mu\nu}(l)\right\rbrack d_{\nu\mu}(l)\nonumber\\
&=&i\frac{\alpha_s}{2\pi}
\int_{-\infty}^{\infty} {dz\over x}\left\lbrack
\theta(z)\theta(x-z)+
 2n(z)\epsilon(z)\right\rbrack
   P_{gg}(z/x)
   \int\frac{dk_{\perp}^2}{k_{\perp}^2} \;\; .
 \end{eqnarray}
As in Eq.(\ref{qselfe2}), 
$\Pi_{g}^{\mu\nu}(l)d_{\nu\mu}(l)/l^2$
can be divided into physical decay and scattering processes
coming from Landau damping at finite temperature.  
This leads to the contribution
 \begin{equation}
 \label{dDgc2}
 \Delta {F}_{h/g}^{(vg)}(y;x)=
   -\frac{\alpha_s}{4\pi}{F}_{h/g}(y;x)
   \int\frac{dk_{\perp}^2}{k_{\perp}^2}
\int_{-\infty}^{\infty} {dz\over x}
  \,\lbrack\theta(z)\theta(x-z)
+2\epsilon(z)n(z)\rbrack
   P_{gg}(z/x)\, .
 \end{equation}
Dividing this integral into different regions, 
as in Eq.~(\ref{eq:cancel}),
one can see that all of the infrared divergences
present in Eq.~(\ref{dDgb2}) are canceled by those in 
Eq.~(\ref{dDgc2}).

An additional virtual correction to the gluon's fragmentation function
is illustrated in Fig.~\ref{fig9}. 
Although this correction contains no 
infrared divergence, it has the normal collinear divergence and thus
will contribute to
the scale dependence of the fragmentation function. This
correction contains the hard part
 \begin{eqnarray}
 \label{hgd}
 {\hat H}_g^{\mu\nu(vq)}(l)&=& \frac{n_f}{2}
   \Bigl[\Pi_{q}^{\mu\sigma}(l)\frac{i}{l^2+i\eta}{+}
   \Pi_{q}^{\mu\sigma*}(l)\frac{-i}{l^2-i\eta}\Bigr]
  {y\over x}\, d_{\sigma\tau}(l)d^{\tau\nu}(l)
   \delta\left({y\over x}{-}\frac{p_h^+}{l^+}\right)
 \nonumber\\
   &=&-n_f {\rm Im}\Bigl[\frac{\Pi^{\mu\sigma}(l)}{l^2+i\eta}\Bigr]
   {y\over x}\, d_{\sigma\tau}(l)d^{\tau\nu}(l)
   \delta\left({y\over x}{-}\frac{p_h^+}{l^+}\right)\;\; .
 \end{eqnarray}
Here, $\Pi_q^{\mu\nu}(l)$ is the quark contribution
to the gluon self-energy.  It has the form
 \begin{equation}
 \label{qloop1}
    \Pi_q^{\mu\sigma}(l)=g_s^2 T_F \int\frac{d^4 k}{(2\pi)^4}
    {\rm Tr}\left[\gamma^{\mu}S(k)\gamma^{\sigma}S(k-l)\right]\, ,
 \end{equation}
Standard manipulations lead to a collinear divergent term
 \begin{eqnarray}
 \label{qloop2}
  &&\lim_{l\rightarrow l_c}
{1\over l^2}\;d_{\nu\mu}(l)\left\lbrack\Pi_q^{\mu\sigma}(l)
\right\rbrack
    d_{\sigma\tau}(l)d^{\tau\nu}(l)
    =\lim_{l\rightarrow l_c}{1\over l^2}\;\left\lbrack
\Pi_q^{\mu\sigma}(l)d_{\sigma\mu}(l)\right\rbrack
 \nonumber\\
  &&\quad=i\frac{\alpha_s}{\pi}
\int_{-\infty}^{\infty} {dz\over x}
 \lbrack \theta(z)\theta(x-z)-2f(z)
\epsilon(z)\rbrack
   P_{gq}(z/x)
   \int\frac{dk_{\perp}^2}{k_{\perp}^2}\;\; ,
 \end{eqnarray}
which gives rise to the virtual correction
 \begin{eqnarray}
 \label{dDgd2}
\Delta {F}_{h/g}^{(vq)}(y;x)=
   -\frac{\alpha_s n_f}{2\pi}{F}_{h/g}(y;x)
   \int\frac{dk_{\perp}^2}{k_{\perp}^2}
   \int_{-\infty}^{\infty} {dz\over x}
   \lbrack\theta(z)\theta(x-z)-
2\epsilon(z)f(z)\rbrack
   P_{gq}(z/x)\;\; .
 \end{eqnarray}
This correction does not contain an infrared divergence,
as mentioned earlier, but the presence of a collinear divergence
implies that it will still contribute to the 
scale dependence of the fragmentation function.

Putting all of these corrections together, we arrive at the
expression
 \begin{eqnarray}
 \label{dDg}
{F}_{h/g}(y;x,Q^2)
  -{F}_{h/g}(y;x,\mu^2)&=&
  \Delta {F}_{h/g}^{(g)}(y;x,Q^2){+}
  \Delta {F}_{h/g}^{(q)}(y;x,Q^2){+}
  \Delta {F}_{h/g}^{(vg)}(y;x,Q^2){+}
  \Delta {F}_{h/g}^{(vq)}(y;x,Q^2)
 \nonumber\\
   &=&\frac{\alpha_s(\mu^2)}{2\pi}
\int_{\mu^2}^{Q^2}\frac{dk_{\perp}^2}{k_{\perp}^2}
   \int_0^{\infty}\frac{dz}{z}
   \left\lbrack{\widetilde \gamma}_{gq}(z;x)
F_{h/s}(y;z,\mu^2)+
   {\widetilde \gamma}_{gg}(z;x)
   {F}_{h/g}(y;z,\mu^2)\right\rbrack
 \end{eqnarray}
for the renormalized gluon fragmentation function.
The modified kernels are
 \begin{mathletters}
 \label{modisplitaa}
 \begin{eqnarray}
 {\widetilde \gamma}_{gq}(z;x)&=&
P_{gq}(z/x)\epsilon(x-z)\lbrack \theta(x-z)-f(x-z)\rbrack
 \label{modisplitaa1}\\
{\widetilde \gamma}_{gg}(z;x)&=&
P_{gg}(z/x)\epsilon(x-z)\lbrack
\theta(x-z)+n(x-z)\rbrack\nonumber\\
&&\qquad-{1\over 2}\delta(1-x/z)\int_{-\infty}^\infty
{d\omega\over x}
\left\lbrace P_{gg}(\omega/x)\epsilon(x-\omega)\lbrack
\theta(\omega)\theta(x-\omega)+
2n(x-\omega)\rbrack\right.\\&&\qquad\qquad\left.
+2n_fP_{gq}(\omega/x)\epsilon(\omega)\lbrack
\theta(\omega)\theta(x-\omega)-2f(\omega)\rbrack\right\rbrace\;\; , \nonumber
 \label{modisplitaa2}
 \end{eqnarray}
 \end{mathletters}
where the splitting functions $P_{ga}(z_f)$ 
have been given in Eqs.(\ref{Pgqq})
and (\ref{Pggg}).

\section{Evolution Equations and Energy Loss in a Thermal Medium}
\label{sec4}

The first order corrections to the
fragmentation functions derived above
can be used to resum the leading powers
of $\alpha_s(\mu^2)\log(Q^2/\mu^2)$ 
in our perturbative expansion.
In order to obtain these leading logarithms,
the transverse momenta of the emitted and 
absorbed thermal partons must be ordered 
from largest to smallest as the quark propagates.
Essentially, this fact arises from the form of the 
nested loop integrals.  In order to obtain a 
double logarithm from
\begin{equation}
\int^{Q^2}{dk_\perp^2\over k_\perp^2}
\int^{Q^2}{dl_\perp^2\over l_\perp^2+k_\perp^2}\;\; ,
\end{equation}
we must have $l_\perp^2>k_\perp^2$.
At zero temperature, this ordering implies that the 
virtuality of the leading parton is similarly ordered.
Absorption processes present in thermal media
disallow this virtuality ordering, so only
the minimum requirement of transverse-momentum 
ordering is observed by our corrections.

As in Eq.(\ref{DGLAP0}), the result of the
resummation can be expressed in the compact form 
of an evolution equation:
 \begin{mathletters}
 \label{Tevolution}
 \begin{eqnarray}
  Q^2\,{d\over dQ^2} {F}_{h/q}(y;x,Q^2)
   &=&\frac{\alpha_s(Q^2)}{2\pi}\int_0^{\infty}\frac{dz}{z}
   \left\lbrack{\widetilde \gamma}_{qq}(z;x)
   {F}_{h/q}(y;z,Q^2)+
   {\widetilde \gamma}_{qg}(z;x)
   {F}_{h/g}(y;z,Q^2)\right\rbrack
 \label{Tevolution1}\\
   Q^2\,{d\over dQ^2} {F}_{h/g}(y;x,Q^2)
   &=&\frac{\alpha_s(Q^2)}{2\pi}\int_0^{\infty}\frac{dz}{z}
  \left\lbrack{\widetilde \gamma}_{gq}(z;x)
F_{h/s}(y;z,Q^2)+
   {\widetilde \gamma}_{gg}(z;x)
   {F}_{h/g}(y;z,Q^2)\right\rbrack\;\; .
 \label{Tevolution2}
 \end{eqnarray}
 \end{mathletters}
These equations express the change in our fragmentation
functions as the probing scale is varied.
At zero temperature, this dependence arises from
the arbitrary distinction between a parton and 
its surrounding vacuum fluctuations.  The theoretical
process in which a quark emits a gluon with 
transverse momentum smaller than the probing scale
cannot be distinguished from the process in which
no such emission occurred.  Hence, this state is considered
part of the vacuum fluctuations inherent in the nature
of the quark's propagation.  On the other hand, if
the transverse momentum between the quark and emitted
gluon is larger than the probing scale, the final state
is physically different from that of a single quark.
In this case, the process represents a physical bremsstrahlung.
Obviously, changes in the probing scale will 
give rise to changes in the effective fragmentation function.

At finite temperature, the probing scale also serves to 
separate the parton from the surrounding thermal bath.
Just as we cannot completely separate a parton 
from its vacuum fluctuations, a parton in
a heat bath cannot be completely separated from the
fluctuations of the bath.  
The idea behind these two
cases is quite similar, but there are some striking 
qualitative differences.  
In particular, 
the vacuum does not carry any energy.  Therefore, 
the total energy of a fragmenting parton is 
conserved under changes of scale:
\begin{equation}
\label{cons}
Q^2\,{d\over dQ^2}\,\sum_h\int dz zD_{q\rightarrow h}(z,Q^2)=0\;\; .
\end{equation}
However, a thermal bath does carry energy.  Hence the 
energy that is attributed to a jet in a thermal medium
will depend on the scale used to probe the jet.

In order to analyze the effect of evolution on 
the fragmentation functions at finite temperature,
we must first resolve the difficulty 
mentioned in Section \ref{sec3a}:
It is not truly consistent
to treat the fragmentation of a low-energy parton
without taking the fragmentation of the bath itself
into account.  To separate the contribution
from distinct partons to our evolution equations
from that of the bath, we 
introduce the `assimilation functions'
$p_{q,g}(z/T)$ and $\bar p_{q,g}(z/T)=1-p_{q,g}(z/T)$.
Phenomenologically,
$p_{q(g)}(z/T)$ represents the probability that a 
quark (gluon) of momentum $z$ will 
thermalize with a bath of temperature $T$ before it
has time to fragment.  Keeping track of only the first 
term in Eq. (\ref{Tevolution1}), we write
\begin{eqnarray}
Q^2{d\over dQ^2}\,F_{h/q}(y;x,Q^2)\,dy/x&=&
Q^2{d\over dQ^2}\,P(x\rightarrow(y,dy);Q^2)\nonumber\\
&=&{\alpha_s(Q^2)\over2\pi}\int_0^\infty {dz\over x}\,
{\widetilde\gamma}_{qq}(z;x)\bar p_q(z/T)
P(z\rightarrow(y,dy);Q^2)\nonumber\\
&&+{\alpha_s(Q^2)\over2\pi}\int_0^\infty {dz\over x}\,
{\widetilde\gamma}_{qq}(z;x)p_q(z/T)
P(z\rightarrow(y,dy);Q^2)\;\; .
\end{eqnarray}
Here, $P(x\rightarrow(y,dy);Q^2)$ is the probability that
a quark of momentum $x$ will decay into a hadron
with momentum between $y$ and $y+dy$ and transverse momentum
$k_\perp^2\leq Q^2$.  In the first term on the 
right-hand side of the second equality, the 
quark maintains its identity after the bremsstrahlung.
Hence the use of $F_{h/q}(y;z,Q^2)$ to describe the 
subsequent fragmentation is appropriate.  In the second
term, however, the quark thermalizes with the medium
before the fragmentation takes place.  Therefore,
it is actually the thermal bath that produces the 
final-state hadron.  

Although the latter contribution naturally appears when calculating
the radiative corrections to the fragmentation functions at finite
temperature as defined in Eq.(\ref{22}),
it is inconsistent with the idea of a distinct parton fragmenting
into an observed hadron while propagating through a thermal bath.  
Experimentally, it belongs in the background and 
should be removed before the data are analyzed.
For this reason, we retain only the first term in our evolution
equation.  This procedure does not come without a price.  In 
particular, the fragmentation function described by 
our evolution equation is no longer rigorously associated
with the thermal matrix element appearing in (\ref{22}).
However, this truncation is necessary in order to 
describe a phenomenologically meaningful function.
Discrepancies between the truncated fragmentation function
described below and that defined in Eq.(\ref{22}) 
are important only when we consider hadrons which
are easily generated by the bath.  Since these 
hadrons are difficult to distinguish experimentally from 
the background, their contribution does not concern us here.

Extending the above analysis to the 
full evolution equation (\ref{Tevolution}),
we arrive at 
\begin{mathletters}
\label{Tev}
\begin{eqnarray}
  Q^2\,{d\over dQ^2} {F}_{h/q}(y;x,Q^2)
   &=&\frac{\alpha_s(Q^2)}{2\pi}\int_0^{\infty}\frac{dz}{z}
   \left\lbrack{\widetilde \gamma}_{qq}(z;x)\bar p_q(z/T)
   {F}_{h/q}(y;z,Q^2)+
   {\widetilde \gamma}_{qg}(z;x)\bar p_g(z/T)
   {F}_{h/g}(y;z,Q^2)\right\rbrack\label{Tev1}\\
   Q^2\,{d\over dQ^2} {F}_{h/g}(y;x,Q^2)
   &=&\frac{\alpha_s(Q^2)}{2\pi}\int_0^{\infty}\frac{dz}{z}
  \left\lbrack{\widetilde \gamma}_{gq}(z;x)\bar p_q(z/T)
F_{h/s}(y;z,Q^2)+
   {\widetilde \gamma}_{gg}(z;x)\bar p_g(z/T)
   {F}_{h/g}(y;z,Q^2)\right\rbrack\;\; .
 \label{Tev2}
\end{eqnarray}
\end{mathletters}
\noindent As explained above, these evolution equations 
contain contributions only from processes in which the 
parton propagating through the thermal medium
remains independent of the medium.

The assimilation
functions, $p_{q,g}(z/T)$, are easy 
to understand phenomenologically,
but difficult to define rigorously.  
They can be determined from phenomenological
models of parton thermalization, such as 
those used in Ref. \cite{pcm}.  All that is required
for Eq. (\ref{Tev}) to be well-defined mathematically 
is that $\bar p_q(\zeta)\rightarrow0$ 
faster than $\zeta$ and $\bar p_g(\zeta)\rightarrow0$ 
faster than $\zeta^3$ as $\zeta\rightarrow0$,
along with the obvious restriction
$0\leq p_{q,g}(\zeta)\leq1$.  In addition,
we impose the physical restriction
$p_{q,g}(\zeta)\rightarrow0$ as $\zeta\rightarrow\infty$.

At this point, we are ready to investigate the 
effects of evolution on simple observables of 
the parton jet.  Defining the jet energies
\begin{mathletters}
\begin{eqnarray}
E_q(x,Q^2)&\equiv&\sum_h\int_0^\infty {y\over x}\,{dy\over x}\,
F_{h/q}(y;x,Q^2)\,;\\
E_g(x,Q^2)&\equiv&\sum_h\int_0^\infty {y\over x}\,{dy\over x}\,
F_{h/g}(y;x,Q^2)\;\; ,
\end{eqnarray}
\end{mathletters}
we have 
\begin{mathletters}
\label{nenev}
\begin{eqnarray}
Q^2\,{d\over dQ^2}E_q(x,Q^2)&=&
{\alpha_s(Q^2)\over2\pi}
\int_0^\infty\,{zdz\over x^2}\left\lbrack
{\widetilde\gamma}_{qq}(z;x)\bar p_q(z/T)
E_q(z,Q^2)+{\widetilde\gamma}_{qg}(z;x)\bar p_g(z/T)
E_g(z,Q^2)\right\rbrack\, ;
\label{nenev1}\\
Q^2\,{d\over dQ^2}E_g(x,Q^2)&=&{\alpha_s(Q^2)\over2\pi}
\int_0^\infty\,{zdz\over x^2}\left\lbrack
{\widetilde\gamma}_{gq}(z;x)\bar p_q(z/T)
E_s(z,Q^2)+{\widetilde\gamma}_{gg}(z;x)\bar p_g(z/T)
E_g(z,Q^2)\right\rbrack\;\; .
\end{eqnarray}
\end{mathletters}
These functions represent the fraction of the initial
parton energy carried by the final hadron jet.
Given initial conditions, we can solve this set of
equations numerically to obtain the 
scale dependence of the jet energy.  

We can get some idea of the effect of evolution 
at large $x\gg T$
by considering $E_{q,g}(z,Q^2_0)=1$ 
at some scale $Q_0^2$.  This is precisely the 
zero-temperature value of $E_{q,g}$, so it 
should be approximately valid at high energy.
This ansatz leads to 
\begin{mathletters}
\label{ansatz}
\begin{eqnarray}
\left.Q^2{d\over dQ^2}\,E_q(x,Q^2)\right|_{Q^2=Q_0^2}
&=&{\alpha_s(Q_0^2)\over2\pi}\left\lbrack \bar\Gamma^{(2)}_{qq}(x)
+\bar\Gamma^{(2)}_{qg}(x)\right\rbrack\label{ansatz1}\\
\left.Q^2{d\over dQ^2}\,E_g(x,Q^2)\right|_{Q^2=Q_0^2}
&=&{\alpha_s(Q_0^2)\over2\pi}\left\lbrack 2n_f\bar\Gamma^{(2)}_{gq}(x)
+\bar\Gamma^{(2)}_{gg}(x)\right\rbrack\;\; ,\label{ansatz2}
\end{eqnarray}
\end{mathletters}
where 
\begin{equation}
\bar\Gamma^{(n)}_{pp'}(x)=\int_0^\infty{dz\over x}
\left({z\over x}\right)^{n-1}{\widetilde\gamma}_{pp'}(z;x)
\bar p_{p'}(z/T)\;\; .
\end{equation}

Contributions to Eq.(\ref{ansatz1}) can be understood in terms
of the energy loss and gain to different partonic degrees
of freedom due to the basic bremsstrahlung/absorption process.
The first and second terms represent energy gain to the 
quark and gluon components of the fragmenting quark, respectively.
Since the vacuum does not carry any energy, energy losses to the 
quark degrees of freedom must be compensated completely
by gains in the gluon sector at zero temperature.  
Explicit calculation yields
\begin{equation}
\bar\Gamma^{(2)}_{qq}(x)|_{T=0}=
-\bar\Gamma^{(2)}_{qg}(x)|_{T=0}=-{4\over3}\,C_F\;\; ,
\end{equation}
in agreement with our expectations.  

As the temperature is increased
from zero, several effects work to alter this
result.  To analyze their contributions
separately, it is useful to replace 
$\bar p_{q,g}(z/T)$ with $1-p_{q,g}(z/T)$
in $\bar\Gamma^{(2)}_{q\,q,g}(x)$.   
The `1' contains the vacuum contributions 
described above, as well as the 
thermal effects of induced emission and 
absorption to $\bar\Gamma^{(2)}_{qq}(x)$ 
and those of Pauli blocking and quark annihilation
to $\bar\Gamma^{(2)}_{qg}(x)$.  
The assimilation functions take the energy flow 
from explicit parton degrees of freedom to the 
heat bath into account.  These contributions 
act to reduce the amount of energy 
available to both quark and gluon degrees 
of freedom in the jet.

Induced gluon emission and absorption  
allow the thermal bath to change explicitly
the energy of the distinct parton degrees of 
freedom.  These processes only involve thermal gluons, so 
they do not contribute to $\bar\Gamma^{(2)}_{qg}(x)$.  
The form of the splitting function $P_{qq}(z_f)$ 
guarantees that the absorption process wins out over that
of induced emission, giving rise to the net fractional 
energy gain 
\begin{equation}
\delta_1\bar\Gamma^{(2)}_{qq}(x)={5\over 6}\,C_F
\left({\pi T\over x}\right)^2+{\cal O}((T/x)^4)
\end{equation}
to the quark degrees of freedom.  
The analogous processes
of Pauli blocking and quark annihilation 
affecting the gluonic degrees of freedom cancel almost exactly,
leaving a net energy gain from annihilation 
that is suppressed by $\exp(-x/T)$.

Our distinct parton degrees of freedom 
lose energy to the medium mainly through the 
direct bremsstrahlung process.  In order to 
obtain an explicit expression for this 
contribution to the energy loss, we
must take a specific form for the assimilation
functions.  For illustrative purposes, we
choose the hard cut-off 
$p_{q,g}(\zeta)=\theta(\eta_{q,g}-\zeta)$, where
$\eta_{q,g}\sim 1$ governs the thermalization
of partons introduced to our thermal medium.
This choice yields the fractional energy losses
\begin{mathletters}
\begin{eqnarray}
\delta_2\bar\Gamma^{(2)}_{qq}(x)&=&-{1\over 2}\,C_F
\left({\eta_q T\over x}\right)^2-{1\over3}\,C_F
\left({\eta_q T\over x}\right)^3+{\cal O}((T/x)^4)\, ; \\
\delta_2\bar\Gamma^{(2)}_{qg}(x)&=&-2C_F\,{\eta_g T\over x}+
C_F\,\left({\eta_g T\over x}\right)^2-{1\over 3}\,C_F\left(
{\eta_g T\over x}\right)^3+{\cal O}(\exp(\eta_g-x/T))\;\; .\label{linear}
\end{eqnarray}
\end{mathletters}
The appearance of a linear (in $T/x$) term in Eq.~(\ref{linear})
is quite striking since it dominates all other corrections 
as $x\rightarrow\infty$.   This term is generated by the
divergent behavior of the soft gluon
emission rate.  Since $P_{qg}(z_f)\rightarrow 2/z_f$ as $z_f\rightarrow0$,
the fractional energy carried by soft gluons saturates.  
This causes the amount of energy absorbed by the bath 
to be proportional to the size of the energy 
window that thermalizes with the bath, which is dictated
by the temperature $T$. Such a linear term will appear in the
asymptotic expansion (as $x \rightarrow\infty$) for any
choice of assimilation function satisfying the
requirements as we have given before. 

Putting all of this together and performing similar
calculations in the gluon sector, we rewrite Eq.(\ref{ansatz}) as
\begin{mathletters}
\label{nansatz}
\begin{eqnarray}
\left.Q^2{d\over dQ^2}\,E_q(x,Q^2)\right|_{Q^2=Q_0^2}
&=&-{\alpha_s(Q_0^2)C_F\over2\pi}\,{T\over x}\left\lbrack 
2\eta_g\left(1-{\eta_gT\over2x}\right)+\eta_q\,{\eta_qT\over2x}
-{5\over 6}\,\pi^2\,{T\over x}+{\cal O}((T/x)^2)
\right\rbrack\label{nansatz1}\\
\left.Q^2{d\over dQ^2}\,E_g(x,Q^2)\right|_{Q^2=Q_0^2}
&=&-{\alpha_s(Q_0^2)\over2\pi}\,{T\over x}
\left\lbrack 2C_A\left(\eta_g-\eta_g\,{\eta_gT\over x}-{1\over3}
\,\pi^2\,{T\over x}\right) \right. \nonumber \\
&+& \left. 2n_fT_F\left(\eta_q\,{\eta_qT\over2x}
-{1\over6}\,\pi^2\,{T\over x}\right)+{\cal O}((T/x)^2)
\right\rbrack\;\; .\label{nansatz2}
\end{eqnarray}
\end{mathletters}
As before, the linear term in the gluon sector is 
generated by the divergence of the 
soft gluon bremsstrahlung rate.  
These expressions indicate that 
the parton energy which 
contributes to a hadronic jet is 
reduced by an amount approximately proportional
to the temperature of the thermal medium as the 
transverse momentum cut-off of the jet is 
increased.  The effect is slightly smaller at lower 
energies due to absorption processes.
This phenomenon arises from the fact that increasing
the momentum cut-off increases the 
amount of transverse phase space available to 
partonic sub-processes, which increases the amount
of energy lost to the bath.  
Although our result was derived from 
a constant input distribution and rather simplistic
assimilation functions, the qualitative effect of evolution is
expected to be similar to that of a more realistic calculation.
In particular, the dominant behavior of the thermalization
mechanism over the phenomena of induced emission,
absorption, annihilation and Pauli blocking
is expected to persist in a complete calculation.

Fluctuations in the thermal medium can also screen the
quark flavor of a jet.  The valence structure 
of a jet is measured by the quark number
\begin{equation}
N_q(x,Q^2)=\sum_{h(q)}\int_0^\infty {dy\over x}\left\lbrack
F_{h/q}(y;x,Q^2)-F_{h/\overline q}(y;x,Q^2)\right\rbrack\;\; ,
\label{qnum}
\end{equation}
where the sum goes over all hadrons $h$
whose valence structure includes $q$.  
Only fragmentation processes in which 
the leading quark contributes to the valence structure
of the observed hadron can contribute to Eq.~(\ref{qnum}),
since all others will cancel in the difference.
In vacuum, $N_q(x,Q^2)$ represents the net flavor of the jet.
Since QCD interactions conserve flavor, 
\begin{equation}
N_q(x,Q^2)|_{T=0}=1
\end{equation}
is independent of 
scale or parton energy.  At finite temperature,
the leading quark in a jet can scatter into the thermal
bath or annihilate with an anti-quark in the bath.
As before, these processes generate a dependence
of the quark number on both scale and parton energy.
Eq.(\ref{Tev1}) leads to the 
evolution equation
\begin{equation}
Q^2\,{d\over dQ^2}N_q(x,Q^2)={\alpha_s(Q^2)\over2\pi}
\int_0^\infty {dz\over x}\,{\widetilde\gamma}_{qq}(z;x)\bar p_q(z/T)
N_q(z,Q^2)
\end{equation}
for $N_q(x,Q^2)$.  

As before, we illustrate the effect of this
evolution equation at high energy by considering $N_q(x,Q^2_0)=1$
at the scale $Q_0^2$.  This gives 
\begin{equation}
\left. Q^2\,{d\over dQ^2}\,N_q(x,Q^2)\right|_{Q^2=Q_0^2}
={\alpha_s(Q_0^2)\over2\pi}\,\bar\Gamma^{(1)}_{qq}(x)\;\; .
\end{equation}
Using the same assimilation function as above, 
we arrive at the expression
\begin{equation}
\bar\Gamma^{(1)}_{qq}(x)=-C_F\,{T\over x}\left\lbrack
\eta_q+\eta_q\,{\eta_qT\over 2x}-{1\over 4}\,\pi^2\,{T\over x}
+{\cal O}((T/x)^2)\right\rbrack\;\; ,
\end{equation}
which indicates a net transfer of flavor from the 
jet to the bach as the momentum scale is increased.
It is interesting to note
that while the effect of the medium on energy loss
becomes larger with the energy of the leading parton,
its effect on flavor loss diminishes as the energy is increased.
This is essentially because the {\it fractional} rather than 
absolute energy loss is comparable to the net flavor loss.

We can also consider the energy loss of 
a quark propagating through a thermal medium
from a slightly different point of view.  
Our interpretation of Eq.(\ref{Tevolution})
indicates that 
\begin{equation}
d{\rm P}_{qq}(x\rightarrow z,Q^2)= 
{\alpha_s(Q^2)\over2\pi}\,
{\widetilde\gamma}_{qq}(z;x)\bar p_q(z/T){dz\over x}\,{dQ^2\over Q^2}
\label{splitprob}
\end{equation}
represents the probability that a leading quark of energy $x$
will split into a quark with energy between $z$ and $(z+dz)$
and transverse momentum between $Q^2$ and $Q^2+dQ^2$
and a gluon. The effective energy loss of the 
leading quark can be summarized as
the expectation value of the emitted gluon's energy.

To obtain a number for this quantity, we must first 
determine a range of values for the probing scale. 
Since we have identified $Q^2$ phenomenologically
with the transverse momentum $k_\perp^2$ 
between the quark and gluon, it is natural
to use the kinematic limit
\begin{equation}
k_\perp^2\leq 4z|x-z|\;\; ,
\end{equation}
as shown in Ref.\cite{ww01}.  This limit
depends on the details of the process, 
so our result cannot be directly related to
a renormalization group analysis.  For this reason, 
the appearance of the running coupling in Eq.(\ref{splitprob})
is not fully consistent with the idea of renormalization.
We will take the coupling evaluated at some fixed
scale, $Q_0^2$, characteristic of our process.  The  
infrared scale $\mu^2$, given as the Debye screening mass, is taken
as a lower limit of the $k_T$ integral.  
These considerations lead to the quark fractional energy loss
\begin{eqnarray}
{\langle z_g\rangle\over x}&=&{\alpha_s(Q_0^2)\over2\pi}
\int_0^\infty {dz\over x}\left(1-{z\over x}\right)\int_{\mu^2}^{4z|x-z|}
{dk_\perp^2\over k_\perp^2}\,
{\widetilde\gamma}_{qq}(z;x)\bar p_q(z/T)\; \nonumber\\
&=&{\alpha_s(Q_0^2)C_F\over2\pi}\left\lbrack
\left( {4\over3}\log{4x^2\over\mu^2}-{49\over 18}\right)
-{\eta_qT\over x}\,\left(\log{4x\eta_qT\over\mu^2}-1\right)
\right.\nonumber\\
&&\qquad\left.-{2\over3}\,\left(\pi T\over x\right)^2
\left(\log{4xT\over\mu^2}+2-\gamma_{E}+{6\zeta'(2)\over\pi^2}
-{3\over4}\,{\eta_q^2\over\pi^2}\right)
+{\cal O}((T/x)^3)\right\rbrack\;\; ,
\end{eqnarray}
where we have again used the hard cut-off form of $\bar p_q(z/T)$.

The first term in this expression represents 
quark energy loss due to vacuum bremsstrahlung,
and is independent of the medium.  Some of the energy lost to 
the quark through these processes re-appears in the 
gluon fragmentation, while the rest is lost
to the medium.  The difference is unimportant here
since we do not track the gluon degrees of freedom.
The second term represents the shift in net quark number 
of the jet as the scale varies.  Its sign indicates that it
represents energy {\it gain} to the quark
when the energy is large.  This is 
because we have considered the normalized energy 
loss to one quark, while the net quark number of the 
jet is actually less than one due to
annihilation effects in the medium.  The result
is that the energy {\it per quark} increases.
The third term corresponds to 
energy that comes from
the medium directly either from 
absorption/emission processes or
via soft quark thermalization.  
This contribution indicates that the absorptive processes in the plasma 
overcome the emissive processes in the absence of 
thermalization ($\eta_q=0$), leading to a net energy gain.
The details of the calculation depend on our choice 
of assimilation function, as well as transverse momentum
cut-offs, but the results are nonetheless interesting.
We note here that secondary scattering effects, 
such as those studied in Ref.~\cite{ww01}, appear only at the 
next-to-leading-log approximation.  These are not taken into
account in our formulae. 

\section{Conclusion}
\label{sec5}

In this paper we have taken a first step in the 
study of parton fragmentation functions in a thermal QCD medium
via thermal field theory. We extended the definition of the
parton fragmentation functions to the case of a thermal medium with finite 
temperature which will reduce to the conventional fragmentation functions
in the zero temperature limit. We then calculated the radiative corrections
to these fragmentation functions within the framework of perturbative QCD
at finite temperature and derived the corresponding DGLAP evolution
equations. The modified evolution equations can be written in the same form
as at zero temperature. However, the effective (or medium modified)
splitting functions depend explicitly on the temperature and
contain terms corresponding to various physical processes such
as radiation, absorption, backward and forward scattering, etc.
The introduction of the temperature causes the effective
splitting functions to depend explicitly on the value of the
initial parton energy. This complicates the renormalization group
analysis of the effective fragmentation functions and causes us to
modifiy their intepretation slightly. However, we have shown that the
modified evolution equations are self-consistent.
There are both linear and logarithmic infrared
divergences in the radiative corrections.  Some cancel among themselves
while others are canceled by the virtual corrections.

As an example of the
application of the evolution equations, we also derived the evolution
equations for the net quark number and the energy loss (gain). 
We find that one has to introduce 'assimilation functions' in order
to separate the hadronization of the thermal bath from that of the
initial parton jet. We find similarly as in an earlier study \cite{ww01}
that the absorption of thermal gluons wins out over 
the stimulated emission, due to the form of $q\rightarrow qg$ splitting
function. This effective medium-induced
energy gain will then reduce the total energy loss experienced by
a propagating parton.

As we have emphasized in the Introduction, we have considered only the
first order corrections in the leading logarithmic approximation. In
this case we only included sequential parton emission or absorption.
Collision-induced bremsstrahlung occurs beyond this order of 
perturbation theory. 
The intermediate thermal gluon will no longer 
be on-shell and its propagator will also develop an imaginary part. 
This will result in collision-induced radiation. When the formation
time of the gluon radiation is larger than the collisional rate which is
contained in the imaginary part of the propagator, there will be
LPM suppression of the induced radiation \cite{amy}. We hope to address this
problem in future studies.



{\it Acknowledgements.} --- This work was supported by the
the Director, Office of Energy
Research, Office of High Energy and Nuclear Physics, Division of
Nuclear Physics, and by the Office of Basic Energy Science,
Division of Nuclear Science, of  the U.S. Department of Energy
under Contract No. DE-AC03-76SF00098, and by
National Natural Science Foundation of China under projects
19928511, 10135030 and 10075031.   E.~W. thanks LBNL Nuclear
Theory Group for its hospitality during the completion of this
work. X.-N.~W. thanks A. Majunder for helpful discussions.
E.~W. and X.-N.~W. also thank the Department of Physics, Shandong
University, where part of the work was completed, for its hospitality


\end{document}